\documentclass[12pt,preprint]{emulateapj}
\usepackage{amsmath}
\usepackage{graphicx}

\newcommand{\lo}{\log L_{2500}}
\newcommand{\lx}{\log L_{2~{\rm keV}}}

\newcommand{\Lo}{L_{2500}}
\newcommand{\Fo}{F_{2500}}
\newcommand{\Lx}{L_{2~{\rm keV}}}

\newcommand{\aox}{\alpha_{\rm ox}}

\newcommand{\lbol}{L_{\rm bol}}

\newcommand{\fmin}{F_{\rm min}}
\newcommand{\lmin}{L_{\rm min}}

\newcommand{\Lthr}{L_{\rm thr}}
\newcommand{\ledd}{\lambda_{\rm edd}}

\newcommand{\gammax}{\Gamma_{\rm X}}

\newcommand{\fvar}{F_{\rm var}}
\newcommand{\sfvar}{\sigma(F_{\rm var})}
\newcommand{\deltat}{\Delta t_{\rm sys}}
\newcommand{\sigmax}{\sigma_{\rm X}}
\newcommand{\sigmai}{\sigma_{\rm I}}

\DeclareRobustCommand{\ion}[2]{%
\relax\ifmmode
\ifx\testbx\f@series
{\mathbf{#1\,\mathsc{#2}}}\else
{\mathrm{#1\,\mathsc{#2}}}\fi
\else\textup{#1\,{\mdseries\textsc{#2}}}%
\fi}
\newcommand{\rev}[1]{{ #1}}

\begin{document}

\title{The tight relation between X--ray and ultraviolet luminosity of quasars}
\author{E.~Lusso\altaffilmark{1} and G.~Risaliti\altaffilmark{1}}
\altaffiltext{1}{INAF--Osservatorio Astrofisico di Arcetri, 50125 Florence, Italy.}

\email{lusso@arcetri.astro.it}
 
\begin{abstract}
The observed relation between the soft X--ray and the optical-ultraviolet emission in active galactic nuclei (AGN) is 
non-linear and it is usually parametrized as a dependence between the logarithm of the monochromatic luminosity at 2500 \AA\ and at 2 keV. Previous investigations have found that the dispersion of this relation is rather high ($\sim0.35-0.4$ in log units), which may be caused by measurement \rev{uncertainties}, variability, and intrinsic dispersion due to differences in the AGN physical properties (e.g. different accretion modes).
We show that, once \rev{optically-selected} quasars with homogeneous SED \rev{and X--ray detection} are selected, and dust reddened and/or gas obscured objects are not included, the measured dispersion 
drops to significantly lower values (i.e. $\sim0.21-0.24$ dex). We show that the residual dispersion is due to some extent to variability, and to remaining measurement uncertainties. Therefore, the real physical intrinsic dispersion should be $<0.21$ dex. Such a tight relation, valid over 4 decades in luminosity, must be the manifestation of an intrinsic (and universal) physical relation between the disk, emitting the primary radiation, and the hot electron corona emitting X--rays.
\end{abstract}


\section{Introduction}
The distribution of X--ray and optical-UV properties in quasars, and their possible dependencies upon redshift, have been the subject of active investigations for more than 30 years \citep{avnitananbaum79}.

Such studies usually parametrized the relation between the X--ray and optical-UV emission as a dependence between the logarithm of the monochromatic luminosity at 2500 \AA, and the $\aox$ parameter, defined as the slope of a power law connecting the monochromatic luminosity at 2 keV and $\Lo$: $\aox=-0.384\times \log[\Lx/\Lo]$.
A strong correlation between $\aox$ and the optical luminosity at 2500$\text{\AA}$ is found, while $\aox$ is only marginally dependent upon redshift (but see \citealt{bechtold03} for different results). The $\aox$ distributions typically cover the range 1.2, 1.8, with a mean value of about 1.5. A fairly significant correlation, albeit with a large scatter, is also found between $\aox$ and the Eddington ratio ($\ledd$, \citealt{2010A&A...512A..34L}, L10 hereafter; see also \citealt{2009MNRAS.392.1124V}).

The $\aox-\Lo$ relation is the by-product of the non-linear correlation between $\Lx$ and $\Lo$ ($\lx= \gamma\lo + \beta$) with a slope $\gamma$ of $0.5-0.7$ found in both optically and X--ray selected AGN samples \citep{vignali03,strateva05,steffen06,just07,2010A&A...512A..34L,2010ApJ...708.1388Y}. This implies that optically bright AGN emit less X-rays (per unit UV luminosity) than optically faint AGN (but see \citealt{yuansiebertbrink98,lafranca95} for a different interpretation). 

Recently the non-linear correlation between $\Lx$ and $\Lo$ has been employed to reliably compute cosmological parameters such as $\Omega_{\rm M}$ and $\Omega_\Lambda$, and to build the first Hubble diagram for quasars which extends up to $z>6$ \citep{2015ApJ...815...33R}, in excellent agreement with the analogous Hubble diagram for supernovae in the common redshift range (i.e. $z\sim0.01-1.4$). 

Understanding the $\Lo-\Lx$ relation thus provides a first hint about the nature of the energy generation mechanism in AGN, it is a first step towards understanding the structure of the AGN accretion disk and X--ray corona, and it can be used as a cosmological probe. 

Yet, the dispersion along the $\Lo-\Lx$ relation is found to be rather high, typically $\geq0.35-0.4$ in log units, which may be presumably caused by the combination of multiple effects such as variability/not--simultaneous observations, poor optical-UV and X--ray data quality, and quasars (QSOs) which have intrinsically red continua and/or host galaxy contamination. All these factors add noise to the correlation.

Here we analyse the $\Lo-\Lx$ relation using a sample of optically selected AGN in the Sloan Digital Sky Survey seventh data release \citep{2011ApJS..194...45S} with X--ray data from the latest release of the 3XMM serendipitous source catalogue, 3XMM--DR5 \citep{2015arXiv150407051R}. 
Our main aim is to use this sample (much larger and more homogeneous than the previous ones in the literature) to understand the origin of the observed dispersion, and to evaluate the intrinsic dispersion of the $\Lo-\Lx$ relation.


We adopt a concordance flat $\Lambda$-cosmology with $H_0=70\, \rm{km \,s^{-1}\, Mpc^{-1}}$, $\Omega_\mathrm{M}=0.3$, and $\Omega_\Lambda=0.7$
\citep{komatsu09}.

\section{The data}
\label{The data}
Our sample starts with the catalogue of quasar properties presented by \citet{2011ApJS..194...45S}, which contains 105,783 spectroscopically confirmed broad-lined quasars. The SDSS quasar sample is cross-matched with the source catalogue 3XMM--DR5 \citep{2015arXiv150407051R}. 3XMM--DR5 is the third generation catalogue of serendipitous X-ray sources available online and contains 565,962 X--ray source detections (396,910 unique X-ray sources) made public on or before 2013 December 31\footnote{http://xmmssc.irap.omp.eu/Catalogue/3XMM-DR5/3XMM\_DR5.html}. The net sky area covered (taking into account overlaps between observations) is $\sim$877 deg$^2$, for a net exposure time $\geq$1 ksec.

For the matching we have adopted a maximum separation of 3 arcsec to provide optical classification and spectroscopic redshift for all objects\footnote{This value is rather conservative. The lists of counterparts with matching radius of 2.146, 3.035, and 3.439 are 90\%, 99\%, and 99.73\% complete, respectively \citep{2009A&A...493..339W}.}. This yields 4,069 XMM observations (2,605 unique sources\footnote{Among these sources, 57 have multiple detections in the 3XMM--DR5 source catalogue, but only one is within 3 arcsec.}, 601 \rev{of which with multiple observations}). 
\rev{The number of unique matches obtained after shifting all 3XMM-DR5 data by 1 arcmin in declination is zero, meaning that there are no spurious associations among the 2,605 objects.}

To define a reasonably ``clean'' sample we have applied the following quality cuts from the 3XMM--DR5 catalogue: SUM\_FLAG$<$3 (low level of spurious detections), and HIGH\_BACKGROUND$=$0 (low background levels)\footnote{For more details the reader should refer to the 3XMM catalogue user guide at the following website http://xmmssc.irap.omp.eu/Catalogue/3XMM-DR5/3XMM-DR5\_Catalogue\_User\_Guide.html.}.
We have also excluded all QSOs in the SDSS-DR7 catalogue flagged as broad absorption line (BAL), and radio emitters with radio loudness (flagged by R\_6CM\_2500A) higher than 10 \citep{kellermann89}. We have further neglected 3 quasars which were included in the BAL quasar catalogue by \citet{2009ApJ...692..758G}, and the moderately radio-loud quasar J001115.23+144601.8, which was not flagged as such in the SDSS catalogue. 
The remaining sample after these cuts is composed by 3,304 XMM observations (\rev{2,155 unique quasars}, 470 \rev{of which with 2 or more observations}).  
\rev{For sources with multiple observations (i.e. 1,619 observations for the 470 quasars) we decided to take the one with the longest EPIC exposure. Our aim is to investigate to what degree the scatter on the $\Lo-\Lx$ relation varies once the best possible sample of individual X--ray detections is taken.  
The choice of the longest X--ray exposure is thus the most appropriate in order to minimise the possible``Eddington bias" due to the flux limit of each observation. We will further examine this point in Section~\ref{X--ray variability}.} 

\rev{We have estimated upper limits on the fluxes for the SDSS quasars that have been pointed by XMM--{\it Newton} but were not detected. We first matched the SDSS-DR7 quasar catalogue with the list of 7,781 observations included in the 3XMM-DR5 catalogue finding 3,481 objects within a circle of 15 arcminutes (half of the field of view of the EPIC cameras). 
We then matched the 3,481 quasars with the list of 2,605 X--ray detected sources finding 2,511 matches. 
This yields a sample of 970 SDSS quasars ($\sim$28\%, 970/3,481) which are in at least one XMM pointing but without a detection. 
Not all the X-ray detected quasars are retrieved with our the adopted circle of 30 arcmin diameter (i.e. 2,511/2,605$\sim$95\%). This is due to the shape of the field of view of the two EPIC cameras that is not circular, meaning that there are regions where quasars are actually pointed, but lie at the edges of the field of view outside our adopted matching area. Additionally, the PN camera is shifted by 2 arcmin with respect to the MOS detectors (centred on the respective telescope optical axes), making the whole field of view more like an ellipses\footnote{To recover the whole X--ray quasar sample of 2,605 quasars we should use something like 17 arcmin. With such radius, we find that the number of undetected quasars is almost 45\%. This is clearly an upper limit on the number of undetected objects since many of them are outside the XMM field of view.}. 

We then use FLIX to compute robust $5\sigma$ (corresponding to a likelihood threshold of 15.1) left censored data points (``upper limits''). 
Radio loud and BAL quasars are neglected following the same approach as above. This yields 789 SDSS quasars.
For 254 quasars FLIX did not find any X--ray match in the soft band, and thus soft X--ray flux values were not available. This leads to a final sample of 535 quasars with X--ray upper limits.} 

To perform our analysis we utilised the observed continuum flux density values at rest-frame 2500 \AA\ ($\Fo$) as compiled by \citet{2011ApJS..194...45S}, which take into account emission line contribution\footnote{These observed flux densities are divided by $(1+z)$ to shift these values into the rest-frame.}. The interested reader should refer to their Section~3 for details on their spectral fitting procedure. Five quasars (2 objects with X--ray detection and 3 upper limits) do not have the $\Fo$ information and have been then excluded. Since uncertainties on $\Fo$ were not provided, we assumed a 2\% \rev{uncertainty} on the continuum flux measurement. The average \rev{uncertainty} on the bolometric luminosity ($\lbol$) estimates in the SDSS quasar catalogue is $\sim$3\%. Given that the bolometric measurements have an additional uncertainty due to the bolometric correction employed, we considered 2\% a reasonable value for the uncertainties on continuum fluxes.

\begin{figure}
   \centering
   \includegraphics[width=\hsize]{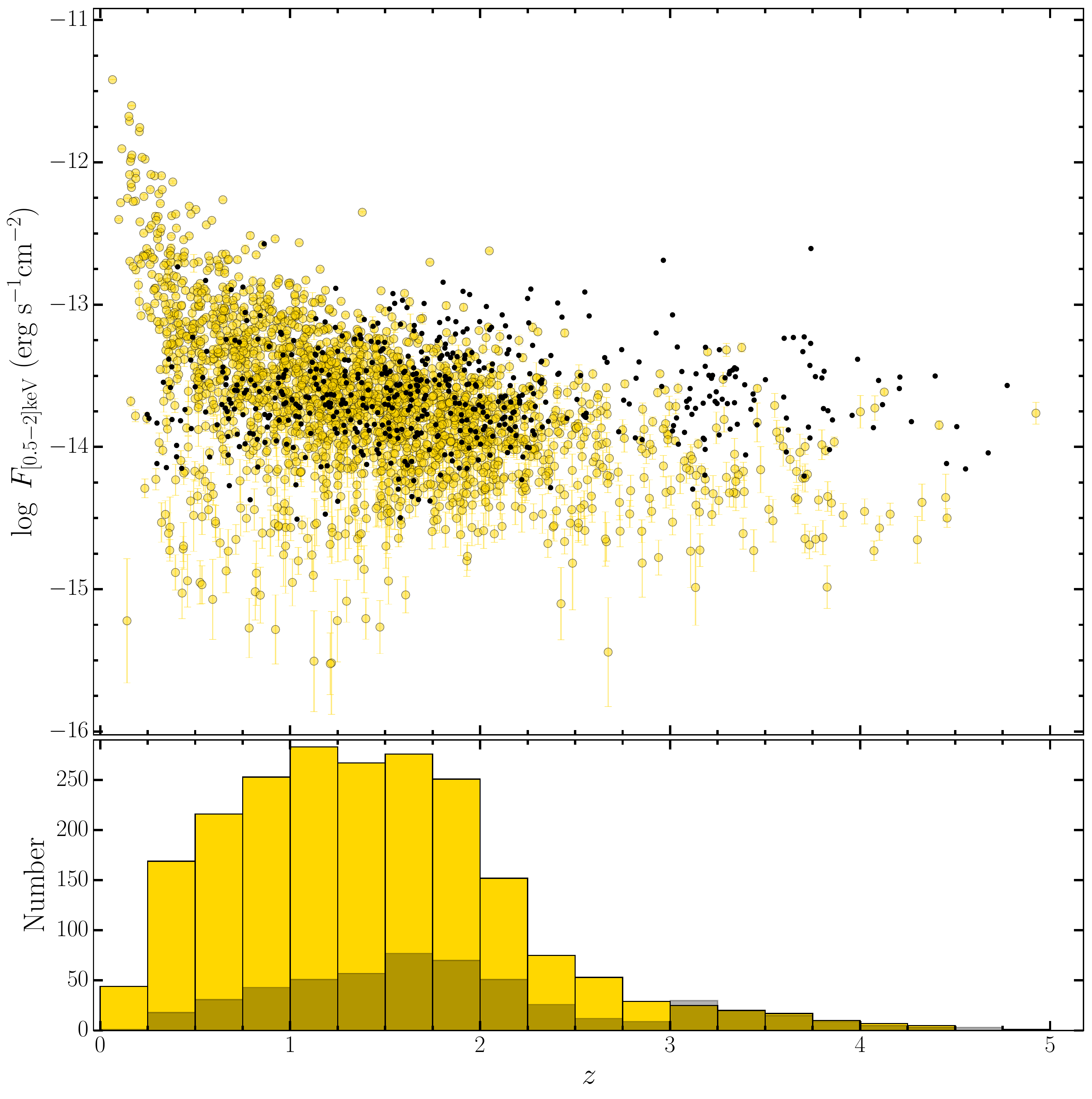}
      \caption{Top panel: Distribution of the observed $[0.5-2]$keV flux versus redshift for the main quasar sample. Upper limits are shown as small black points. 
      Bottom panel: Redshift distribution for the main quasar sample (yellow histogram) and for censored data (light black histogram).}
   \label{fluxSz}
   \end{figure}

The main sample considered in the following analysis is shown in Figure \ref{fluxSz} and it is composed by \rev{2,685} quasars (2,153 X--ray detections and \rev{532} upper limits) spanning a redshift range of 0.065--4.925. 

The relevant source properties of the \rev{main X--ray detected and censored quasar samples are reported in Tables~\ref{sampledet} and ~\ref{sampleul}, respectively}. 
\begin{table*}
\centering
\caption{Optical and X--ray properties of the main X--ray detected quasar sample. \label{sampledet}}
\begin{tabular}{ccccccccccc}
\hline\hline\noalign{\smallskip}
 SDSS Name & RA & DEC  & $z^{\mathrm{a}}$ & $\lo^{\mathrm{b}}$ &  $\lx^{\mathrm{c}}$  &  DETID$^{\mathrm{d}}$  & $\Gamma_1^{\mathrm{e}}$ & $\Gamma_2^{\mathrm{f}}$  & $\Gamma_{\rm X}^{\mathrm{g}}$ & S/N$^{\mathrm{h}}$  \\
           &   J2000.0   &  J2000.0  &     &    &  & & &  & &  \\
\noalign{\smallskip}\hline\noalign{\smallskip}
 000355.49+000736.4 & 0.98121 & 0.126804 & 1.028 & 29.98 & 26.46 $\pm$0.03 & 103057510010006 & 0.78 & 0.62 & 1.67 & 16.90 \\
 000439.97-000146.4 & 1.16656 &  -0.029582 &  0.583 & 29.23 & 25.02$\pm$ 0.12 & 103057510010044 & -0.66 &  -1.05 & 2.17 & 3.98 \\
 000456.17+000645.5 & 1.23405 & 0.112644 & 1.040 & 29.87 &   26.60$\pm$ 0.03 & 103057510010004 & 1.05 & -0.26 & 1.53 & 19.09 \\
\hline\noalign{\smallskip}                      
\end{tabular}                                   
\flushleft\begin{list}{}{Notes---This table is presented in its entirety in the electronic edition; a portion is shown here for guidance.}
\item[$^{\mathrm{a}}$]Spectroscopic redshifts from the SDSS-DR7 quasar catalogue (Z\_HW, improved redshifts from \citealt{2010MNRAS.405.2302H}).  
\item[$^{\mathrm{b}}$]Monochromatic 2500\AA\ luminosities $\rm{(erg\,s^{-1}Hz^{-1})}$ from the SDSS-DR7 quasar catalog. 
\item[$^{\mathrm{c}}$]Monochromatic 2~keV luminosities $\rm{(erg\,s^{-1}Hz^{-1})}$ of the $1^{\rm st}$ XMM longest exposure. 
\item[$^{\mathrm{d}}$] Number which identifies each entry (unique to each X--ray detection) in the 3XMM-DR5 catalog.
\item[$^{\mathrm{e}}$] Slope estimated from the SDSS photometry in the $0.3-1\mu$m range.
\item[$^{\mathrm{f}}$] Slope estimated from the SDSS photometry in the $1450-3000$\AA\ range.
\item[$^{\mathrm{g}}$]X--ray photon index of the $1^{\rm st}$ XMM longest exposure estimated from the slope between the luminosities at 1 and 5 keV. These $\Gamma_{\rm X}$ values have been used to computed $\Lx$ and to select the clean quasar sample.
\item[$^{\mathrm{h}}$] X--ray signal-to-noise of the $1^{\rm st}$ XMM longest exposure: S/N=EP\_8\_CTS/EP\_8\_CTS\_ERR.
\end{list}                                      
\end{table*}                                                                    
\begin{table*}
\centering
\caption{Optical and X--ray properties of the main X--ray censored quasar sample. \label{sampleul}}
\begin{tabular}{cccccccc}
\hline\hline\noalign{\smallskip}
 SDSS Name & RA & DEC  & $z^{\mathrm{a}}$ & $\lo^{\mathrm{b}}$ &  $\lx^{\mathrm{c}}$ & $\Gamma_1^{\mathrm{d}}$ & $\Gamma_2^{\mathrm{e}}$  \\
           &   J2000.0   &  J2000.0  &     &    &  & &   \\
\noalign{\smallskip}\hline\noalign{\smallskip}
000525.12+001745.2 & 1.35469 & 0.295899 & 1.030 & 30.13 &         26.13$\pm$ 0.08 & 1.29 & 0.48\\
001030.55+010006.0 & 2.62731 & 1.00168   & 0.378 & 29.38 &         25.11$\pm$ 0.06 & 0.33 & -0.53\\
001201.87+005259.7 & 3.00782 & 0.883254 & 1.637 & 30.29 &         26.51$\pm$ 0.10 & 0.60 & -0.78\\
\hline\noalign{\smallskip}                      
\end{tabular}                                   
\flushleft\begin{list}{}{Notes---This table is presented in its entirety in the electronic edition; a portion is shown here for guidance.}
\item[$^{\mathrm{a}}$]Spectroscopic redshifts from the SDSS-DR7 quasar catalogue (Z\_HW, improved redshifts from \citealt{2010MNRAS.405.2302H}).  
\item[$^{\mathrm{b}}$]Monochromatic 2500\AA\ luminosities $\rm{(erg\,s^{-1}Hz^{-1})}$ from the SDSS-DR7 quasar catalog. 
\item[$^{\mathrm{c}}$]Monochromatic 2~keV luminosities $\rm{(erg\,s^{-1}Hz^{-1})}$ of the $1^{\rm st}$ XMM longest exposure. 
\item[$^{\mathrm{d}}$] Slope estimated from the SDSS photometry in the $0.3-1\mu$m range.
\item[$^{\mathrm{e}}$] Slope estimated from the SDSS photometry in the $1450-3000$\AA\ range.
\end{list}                                      
\end{table*}                                                                    

\subsection{X--ray luminosities}
\label{X--ray luminosities}
To compute the rest-frame monochromatic luminosities at 2 keV we first estimated the fluxes in the soft (0.5--2 keV) and hard (2--12 keV) energy bands as the sum of the EPIC fluxes listed in the 3XMM--DR5 catalogue in bands 2 and 3, and bands 4 and 5, respectively. Uncertainties on these fluxes are computed by summing in quadrature the catalogued flux \rev{uncertainties} in bands 2 and 3 and in bands 4 and 5 for the soft and the hard band, respectively.  
A power-law spectral model with a photon index, $\Gamma_{\rm X}=1.7$ and a hydrogen absorbing column density of $N_{\rm H} = 3 \times 10^{20}$ cm$^{-2}$ was assumed to convert count rates into fluxes \citep{2015arXiv150407051R}.

To plot the X--ray information in the $\log \nu-\log\nu L(\nu)$ rest-frame plane we have first estimated the luminosity at the geometric mean of the soft (1 keV) and hard (5 keV) energy bands by assuming an average photon index of 1.7 in both bands. Luminosities are then blueshifted to the rest-frame. The rest-frame monochromatic luminosities at 2 keV is finally obtained by interpolation if the source redshift is lower than 1, by extrapolation considering the slope between the luminosities described above for higher redshifts. Uncertainties on monochromatic luminosities ($L_\nu\propto \nu^{-\gamma}$) from interpolation (extrapolation) between two values $L_1$ and $L_2$ are computed as 
\begin{equation}
\label{uncertainties}
\delta L = \sqrt{\left( \frac{\partial L}{\partial L_1}\right)^2 (\delta L_1)^2 + \left( \frac{\partial L}{\partial L_2}\right)^2 (\delta L_2)^2}.
\end{equation}  

For X--ray undetected quasars we have estimated the EPIC fluxes in bands 2 and 3 as the mean of the upper limit flux values in all cameras.
The soft (0.5--2 keV) flux is then the sum of bands 2 and 3
\begin{equation}
F_{[0.5-2]\rm keV, ul} = \frac{\sum_b b\_2\_FUPL}{n} + \frac{\sum_b b\_3\_FUPL}{n},
\end{equation}
where $b=$MOS1, MOS2, and pn and $n$ is the number of cameras with non-zero flux value. Monochromatic 2 keV luminosities have been computed using a photon index of 1.7.

For each detected object we have also computed the EPIC sensitivity ($5\sigma$ minimum detectable flux) at 2 keV. To do that, we have considered the pn, MOS1, and MOS2 on-time\footnote{The total good exposure time (in seconds) of the CCD where the source is detected.} and off-axis values, where both MOS1 and MOS2 on-time and off-axis have been combined. The total MOS on-time and off-axis are the largest and smaller values of the two individual cameras, respectively. 
We then estimated the minimum detectable flux in the soft band as a function of the exposure time following the relations plotted in Figure~3 by \citet{2001A&A...365L..51W} for both pn and MOS. We then corrected this sensitivity for the pn and MOS vignetting factor as a function of their respective off-axis values. 
The same vignetting correction for both pn and MOS has been considered. 
The sensitivity flux values at 2 keV ($\fmin$) are then estimated assuming a photon index of 1.7 and finally combined. We have taken the sum of the pn and MOS fluxes in the case where both values are available. 


\begin{table*}                              
\begin{center}                                 
\caption{Results from correlations analysis of censored data. \label{tbl-2}}
\begin{tabular}{lcccccccc}                           
\hline\hline\noalign{\smallskip}                
Sample & $\gamma$  & $\beta$   & $f$   & $\gamma$ & $\beta$ & $\sigma$ & N$_{\rm ul}$ &  N$_{\rm tot}$ \\
\cline{5-7}\noalign{\smallskip}
  \multicolumn{9}{c}{\hspace{4.5cm} LINMIX\_ERR \hspace{4.3cm} EM} \\
  \cline{2-4}\noalign{\smallskip}
 Main                    & 0.582$\pm$0.014  & 8.664$^{+0.403}_{-0.417}$  & 0.162$\pm$0.005  & 0.586$\pm$0.014    & 8.523$\pm$0.428 & 0.43 & 532 & 2685  \\  
 E(B--V)$\leq$0.1 & 0.592$\pm$0.015  & 8.345$^{+0.471}_{-0.461}$  & 0.150$\pm$0.005  & 0.596$\pm$0.016    & 8.237$\pm$0.492 & 0.41 & 421 &  2319  \\ 
 E(B--V)$\leq$0.1 -- S/N$>$3 & 0.593$\pm$0.016  & 8.334$^{+0.472}_{-0.482}$  & 0.150$\pm$0.005  & 0.596$\pm$0.016    & 8.214$\pm$0.494 & 0.42 & 430 &  2319  \\ 
 E(B--V)$\leq$0.1 -- S/N$>$5 & 0.583$\pm$0.017  & 8.585$^{+0.518}_{-0.519}$  & 0.173$\pm$0.004 & 0.584$\pm$0.018 & 8.562$\pm$0.537 & 0.45 &  635 &  2319  \\    
 E(B--V)$\leq$0.1 -- S/N$>$5 -- $1.6\leq\gammax\leq2.8$ & 0.584$\pm$0.015  & 8.627$^{+0.468}_{-0.462}$  & 0.113$\pm$0.005 & 0.583$\pm$0.016 & 8.658$\pm$0.476 & 0.35 & 527 &  1826  \\   
 
 E(B--V)$\leq$0.1 -- S/N$>$5 -- $1.9\leq\gammax\leq2.8$ & 0.618$\pm$0.019  & 7.570$^{+0.580}_{-0.568}$  & 0.106$\pm$0.006 & 0.616$\pm$0.019 & 7.604$\pm$0.589 & 0.34 & 485 &  1228  \\   
\hline\noalign{\smallskip}        
\end{tabular}                                   
\end{center}                                   
\end{table*}                                                                       
   \begin{figure}
   \centering
   \includegraphics[width=\hsize]{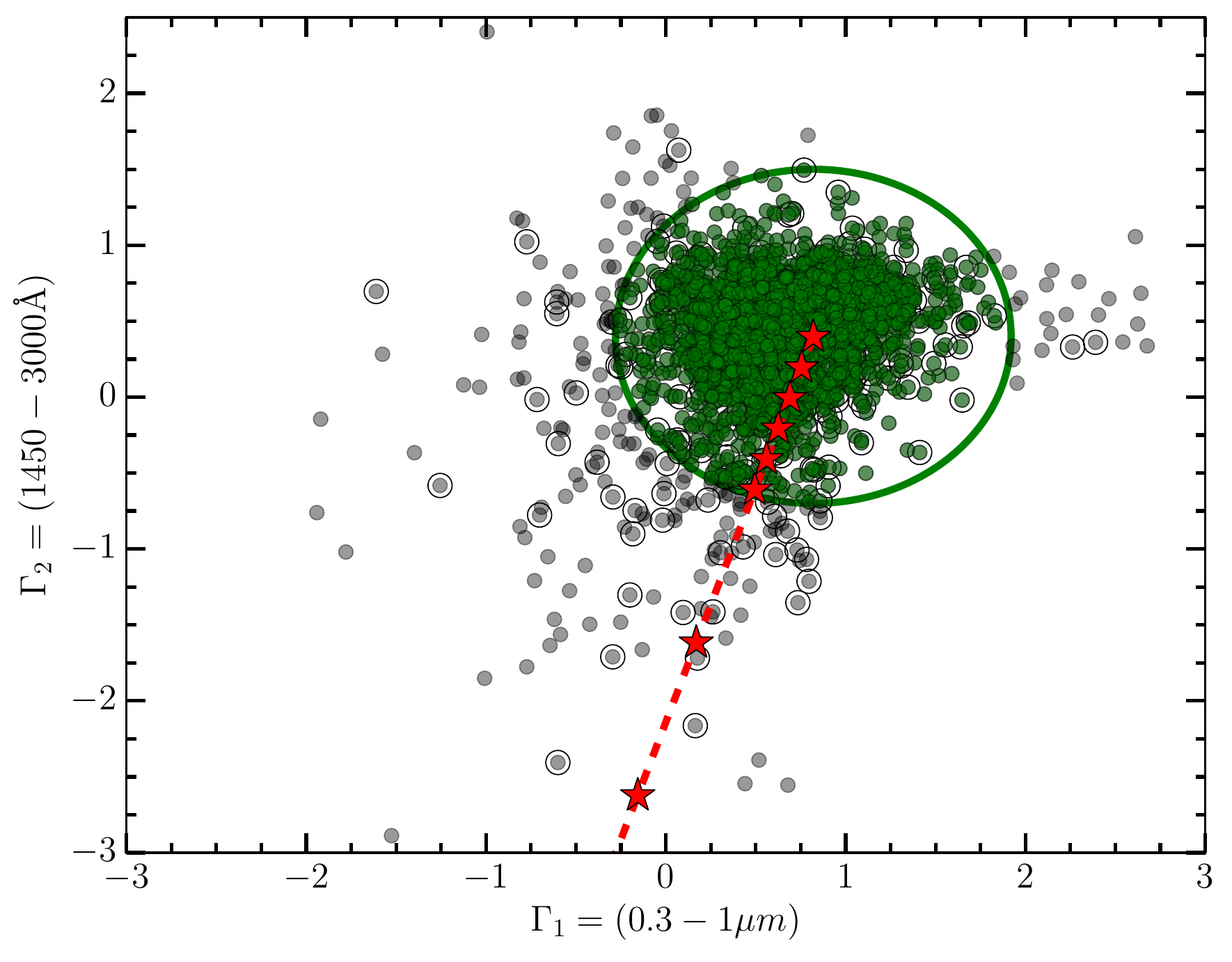}
      \caption{Distribution of the whole quasar sample (excluding radio-loud and BAL sources) in a $\Gamma_{1}-\Gamma_{2}$ plot, where $\Gamma_{1}$ and $\Gamma_{2}$ are the slopes of a power law in the $\log\nu-\log(\nu L_\nu)$ plane, at 0.3--1~$\mu$m and 1450--3000~\AA, respectively. The dashed red line is obtained by assuming increasing dust extinction following the extinction law of \citet{prevot84} for the quasar SED estimated by \citet{2006AJ....131.2766R}. The red stars represent the value of $\Gamma_{1}-\Gamma_{2}$ for $E(B-V)=(0.0,0.02,0.04,0.06,0.08,0.1,0.2,0.3)$. The green circle is centred at $E(B-V)=0.0$ with a radius of 1. ($E(B-V)\simeq0.1$). Sources within the circle are highlighted in green. Open circles highlight X--ray non-detections.}
   \label{redquasars}
   \end{figure}

\section{Statistical analysis}
\label{Statistical analysis}

Previous studies on optically selected AGN reported a relationship between $\Lx$ and $\Lo$ in the form $\Lx \propto \Lo^\gamma$. The best-fit values of the exponent are between $0.7 \div 0.8$ \citep{avnitananbaum82,chanan83,krisscanizares85,wilkes94,yuansiebertbrink98,vignali03,strateva05,steffen06,just07,2010A&A...512A..34L,2010ApJ...708.1388Y}, estimated from an ordinary least-square (OLS) bisector analysis (\citealt{isobe90}). The bisector treats X and Y variables symmetrically, and it has been usually justified by the fact that (1) the choice of the independent variable (between $\Lo$ and $\Lx$) was not straightforward, and (2) methods which minimize residuals of the dependent variable are subject to effects caused by the large observed luminosity dispersion (i.e. 0.35--0.4 dex, see \citealt{2007MNRAS.377.1113T}). 

The observed $\Lo-\Lx$ relation (or its by-product: the $\aox-\Lo$ relation) provides insights into the radiation mechanism in quasars. The non-linearity of such correlation implies that more optically luminous AGN emit less X--rays per unit of UV luminosity than less luminous AGN.
The X--ray properties in quasars have been attributed to the presence of a plasma of relativistic electrons at high temperatures ($T\sim10^{7-8}$K, the so-called {\it corona}) in the vicinity of the accreting supermassive black hole. 
Optical--UV photons from the accretion disk (parametrised by $\Lo$) are Compton up-scattered by hot electrons and lead to the formation of a power law spectrum in the X--rays (parametrised by $\Lx$) accompanied by a high energy cut-off at the electrons' temperature (\citealt{1991ApJ...380L..51H, 1993ApJ...413..507H,2008A&A...485..417D}).
The observed correlation thus suggests that disk and corona are coupled, and that the corona parameters should be then dependent on the UV luminosity. 
In this framework, $\Lo$ and $\Lx$ are not independent parameters, and therefore the regression methods to fit the data needs to be chosen carefully. 

The situation is even more complex in the case of samples with censored data.  
We adopted the LINMIX\_ERR\footnote{This algorithm has been implemented in Python and its description can be found at http://linmix.readthedocs.org/en/latest/src/linmix.html.} method \citep{2007ApJ...665.1489K}, which is argued to be among the most robust regression algorithms with the possibility of reliable estimation of intrinsic random scatter on the regression ($f$). 
LINMIX\_ERR accounts for measurement \rev{uncertainties} on both independent and dependent variable, nondetections, and intrinsic scatter by adopting a Bayesian approach to compute the posterior probability distribution of parameters, given observed data.

To investigate whether the fitting results depend on the adopted LINMIX\_ERR method, we have also considered the Astronomy Survival Analysis software package (ASURV rev. 1.2; \citealt{isobe90}; \citealt{lavalley92}), which is widely used in the literature. ASURV implements the bivariate data-analysis methods and also properly treats censored data using the survival analysis methods (\citealt{feigelsonnelson85}; \citealt{isobe86}). We have employed the full parametric estimate and maximized (EM) regression algorithm to perform the linear regression of the data, and the semiparametric Buckley-James regression algorithm (Buckley \& James 1979). 
The EM regression algorithm is based on the ordinary least-squares regression of the dependent variable Y against the independent variable X (OLS[Y$|$X]). The regression line is defined in such a way that it minimizes the sum of the squares of the Y residuals. However, this regression method is less powerful than LINMIX\_ERR since it does not account for measurement \rev{uncertainties}, and it does not provide an estimate of the intrinsic scatter. 
For censored data, we will present our findings in Section~\ref{The lolx relation: censored data} from both the LINMIX\_ERR and EM regression methods, although the latter is reported just for comparison with previous works in the literature (in all cases the results from the Buckley-James regression algorithm agreed with EM within the uncertainties).

Censored sample are likely to be unbiased, but the analysis of the scatter along the $\Lo-\Lx$ relationship may not be straightforward, since it strongly depends on the weights assumed in the fitting algorithm. Additionally, there is often the situation in surveys where upper limits are not provided. 
In the case of flux-- (or magnitude--) limited surveys, objects with an expected luminosity (based on the observed $\Lo-\Lx$ relation) near to the sample flux limit will be observed only in case of positive fluctuations. Considering only detections may thus introduce a bias in the $\Lo-\Lx$ relationship, and this should be more relevant in the X--rays, since the relative flux/luminosity interval is much smaller than in the optical--UV. 
Therefore, one needs to find an alternative method to obtain a sample where biases are minimised even without the inclusion of censored data.
One possibility is to include only objects that would be observed even in case of negative flux fluctuations. We explored the flux-limit bias in the X--ray detected quasar sample in appendix~\ref{The flux-limit bias}, where we prove that this bias is not significantly affecting our main results. 
We have thus examined the correlation between $\Lo$ and $\Lx$ where nondetections are neglected.

To this goal, we employed an orthogonal distance regression (ODR) fitting procedure
in addition to the LINMIX\_ERR algorithm. The ODR regression treats X and Y variables symmetrically and minimizes both the sum of the squares of the X and Y residuals\footnote{http://docs.scipy.org/doc/scipy/reference/odr.html}. 
We note that the LINMIX\_ERR and ODR algorithms are mathematically different and, in principle, should not be used interchangeably. However, the use of multiple fitting methods, although distinct, is still useful, especially in the case of large scatter. 

\subsection{The $\Lo-\Lx$ relation: censored data}
\label{The lolx relation: censored data}
 
We computed slope and intercept of the $\Lo-\Lx$ relation for the main sample and we investigated how the fit parameters vary depending on possible selection criteria. The findings from the EM regression and the LINMIX\_ERR algorithms are summarised in Table~\ref{tbl-2}\footnote{We considered the dispersion value output of the EM regression algorithm as representative of the scatter along the correlation including censored data.}. 

Comparing our best-fit parameters with those obtained from optically selected samples, we find that our slope is \rev{fully consistent within $1\sigma$ with the results presented by S06 ($\beta_{S06}=0.642\pm0.021$) and J07 ($\beta_{J07}=0.636\pm0.018$), and with the ones by L10 ($\beta_{L10}=0.599\pm0.027$) from OLS(Y$|$X)}. 

We note that continuum flux measurements in the Shen et al. catalogue were neither corrected for intrinsic extinction/reddening, nor for host contamination. Therefore, we singled out a sub-sample of objects where both reddening and host contaminations are reduced at minimum.  
To minimise host-galaxy and reddening contamination we followed a similar approach as in \citet{2015ApJ...815...33R}. We computed for each object the slope $\Gamma_{1}$ of a $\log(\nu)-\log(\nu L\nu)$ power law in the 0.3--1 $\mu$m (rest frame) range, and the analogous slope $\Gamma_{2}$ in the 1450--3000~\AA\ range (rest frame).  
The $\Gamma_{1}-\Gamma_{2}$ distribution is shown in Figure~\ref{redquasars}. We selected all sources with $\Gamma_{1}-\Gamma_{2}$ centred at $E(B-V)=0.0$ with a radius of 1.1, which roughly corresponds to $E(B-V)\simeq0.1$. We found \rev{2,319} quasars that matched this criteria (\rev{421} upper limits).

To avoid large uncertainties on the X--ray flux measurements due to unreliable source counts, we have considered as upper limits all X--ray detected quasars with a ratio between the EPIC source count in the 0.2--12.0 keV band and its uncertainty lower than 3 (S/N=EP\_8\_CTS/EP\_8\_CTS\_ERR$\leq$3, i.e. we have considered X--ray detected all objects with at least 3 sigma counts measurement). We also examined a higher threshold of S/N=5.

Since soft X--ray fluxes may contain some level of absorption, we included only X--ray detected quasars with a photon index $\gammax$ in the range 1.6--2.8, which roughly corresponds to an average $\gammax\sim2$ with a dispersion of 0.3. Our ``photometric" $\gammax$ values represent the slope between the 1 and 5 keV luminosities and, although they cannot be considered as reliable as the spectroscopic measurements, are reasonable tracers of X--ray absorption. 
To further minimise the level of X--ray absorption, we have also studied a narrower $\gammax$ interval of 1.9--2.8 ($\langle \gammax \rangle\sim2.1$ with a dispersion of 0.2, consistent with  \citealt{2009ApJS..183...17Y}). Given the observed $\gammax$ range (up to 2.8), some soft-excess contribution for low-z QSOs might be still present. 
We have thus repeated the analysis further reducing the $\gammax$ range (up to 2.4), but, besides loosing statistics, our results are not affected.

The conditions listed above have the only aim of selecting a sample of quasars with homogeneous SEDs and to minimise the number of red/reddened quasars in both optical and X--ray bands.
These requirements yield a sample of \rev{1,228} quasars, \rev{485} of which are upper limits. Fit parameters for such a sample (slope, intercept, and dispersion) are fully consistent with previous estimates in the literature. 
Our exploration of the parameter space shows that the dispersion of the $\Lo-\Lx$ relation is mainly driven by poor X--ray data, X--ray absorption, and quasars with red continua and/or host galaxy contamination for which the optical flux measurement cannot be properly recovered. 

   \begin{figure}
   \centering
   \includegraphics[width=\hsize]{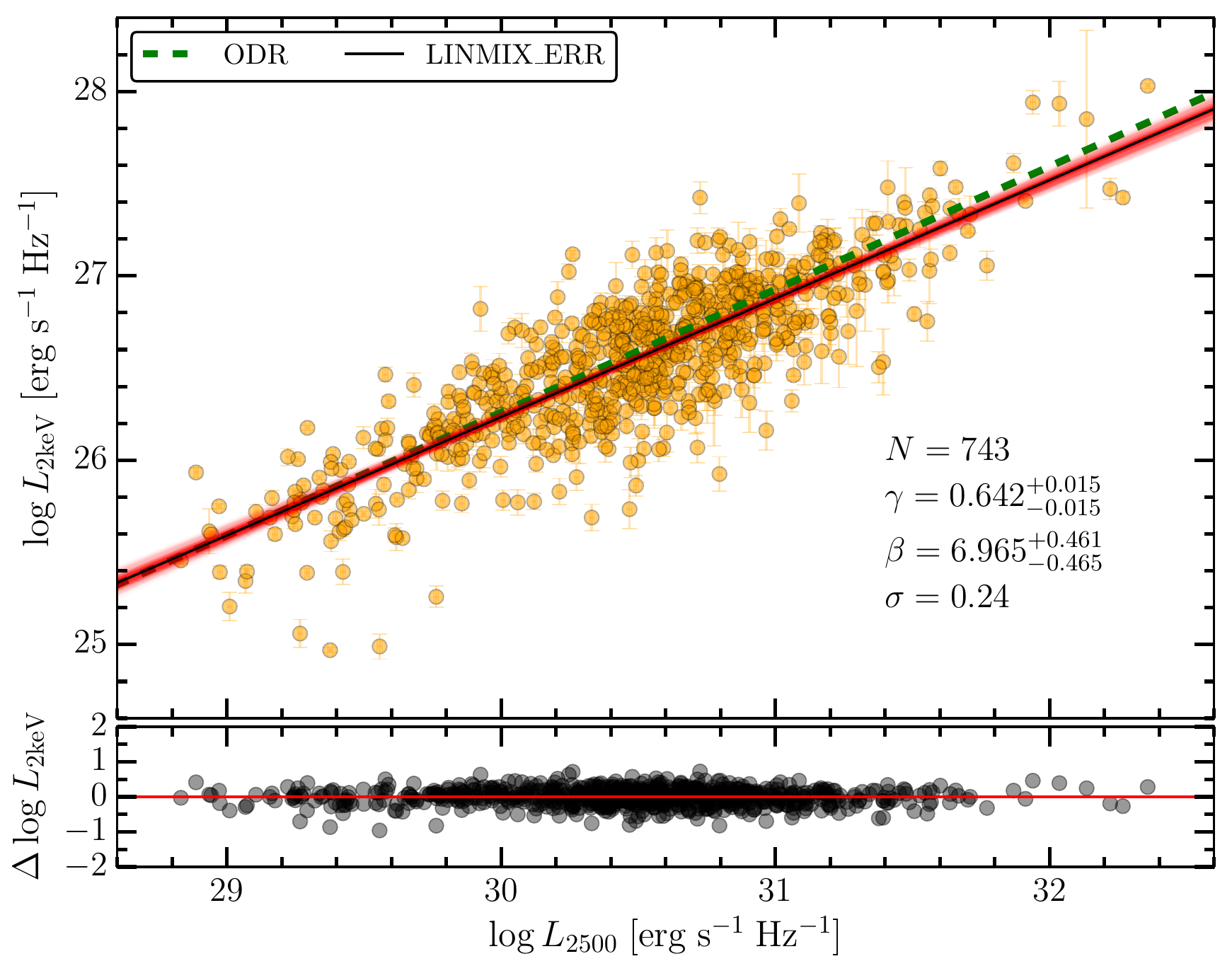}
\caption{Rest-frame monochromatic luminosities $\lx$ against $\lo$ for the X--ray detected (orange circles) quasar samples as described in \S~\ref{The lolx relation (detections)}. The results from the ODR regression (dashed line) and the LINMIX\_ERR (thin solid line) for the selected sample are also reported. Red thin lines represent 400 different realisations of the $\Lo-\Lx$ relation from LINMIX\_ERR. The LINMIX\_ERR regression results for the dispersion, slope, and intercept (with their \rev{uncertainties}) are also reported. The lower panel shows the residuals of $\lx$ and $\lo$ with respect to the LINMIX\_ERR best-fit line.}
         \label{lolx}
   \end{figure}

\subsection{The $\Lo-\Lx$ relation: X--ray detected data}
\label{The lolx relation (detections)}
\begin{table*}                              
\begin{center}                                 
\caption{Results from correlations analysis of detected data. \label{tbl-3}}
\begin{tabular}{lcccccccc}                           
\hline\hline\noalign{\smallskip}                
Sample & $\gamma$  & $\beta$   & $f$ &  $\sigma$  & $\gamma$ & $\beta$ & $\sigma$ &  N$_{\rm tot}$ \\
\cline{6-8}\noalign{\smallskip}
  \multicolumn{9}{c}{\hspace{4.5cm} LINMIX\_ERR \hspace{4.3cm} ODR} \\
  \cline{2-5}\noalign{\smallskip}
 Main                    & 0.583$\pm$0.014 & 8.697$^{+0.415}_{-0.412}$ & 0.147$\pm$0.005 & 0.42 & 0.593$\pm$0.010 & 8.558$\pm$0.303 & 0.45 & 2153  \\    
 E(B--V)$\leq$0.1 & 0.596$\pm$0.016  & 8.279$^{+0.460}_{-0.478}$  & 0.135$\pm$0.005  & 0.40 & 0.618$\pm$0.011 & 7.773$\pm$0.346 & 0.44 &  1898  \\    
 E(B--V)$\leq$0.1 -- S/N$>$3 & 0.596$\pm$0.015  & 8.279$^{+0.462}_{-0.469}$  & 0.135$\pm$0.005  & 0.40 & 0.618$\pm$0.011 & 7.773$\pm$0.346 & 0.44 &  1889  \\   
 E(B--V)$\leq$0.1 -- S/N$>$5 & 0.589$\pm$0.015  & 8.539$^{+0.456}_{-0.465}$  & 0.113$\pm$0.004  & 0.35 & 0.619$\pm$0.012 & 7.752$\pm$0.363 & 0.38 &  1683  \\   
 E(B--V)$\leq$0.1 -- S/N$>$5 -- $1.6\leq\gammax\leq2.8$ &  0.596$\pm$0.014  & 8.392$^{+0.415}_{-0.398}$  & 0.074$\pm$0.003  & 0.28 & 0.634$\pm$0.012 & 7.332$\pm$0.365 & 0.30 &  1298  \\   
 E(B--V)$\leq$0.1 -- S/N$>$5 -- $1.9\leq\gammax\leq2.8$ &  0.642$\pm$0.015  & 6.965$^{+0.461}_{-0.465}$  & 0.053$\pm$0.003  & 0.24 & 0.667$\pm$0.013 & 6.246$\pm$0.378 & 0.24 &  743   \\   
\hline\noalign{\smallskip} 
\end{tabular}                                   
\end{center}                                   
\end{table*}                                                                       

We then repeated the analysis in Section~\ref{The lolx relation: censored data} only for the sample of X--ray detected quasars where we applied the same series of filters as the ones already discussed. The results are plotted in the bottom panel of Figure~\ref{lolx}, while the findings from the regression algorithms for the different selection criteria are presented in Table~\ref{tbl-3}.
The dispersion on this final sample of X--ray detected quasars reduces from $\sim$0.45 to 0.24 dex by applying the same set of filters as the ones of the censored data, which is significantly lower than what previously reported in literature (i.e. $>0.35$ \citealt{vignali03,strateva05,steffen06,just07,2010A&A...512A..34L,2010ApJ...708.1388Y}). Additionally, there is no significant variation on both slope and intercept (within their uncertainties) among the different selections, with the slope being rather constant around 0.6--0.65. The results from the ODR fitting procedure seems to show slightly steeper slopes than LINMIX\_ERR, although the disagreement is below 2$\sigma$. 

We stress that our selection criteria are extremely simple and they have the only aim to favour blue quasars having homogeneous SED, where we can robustly estimate both $\Lo$ and $\Lx$ with minimum contamination from host-galaxy/reddening, and absorption in X--rays. The observed slope and intercept estimated for this sample can be considered representative of {\it intrinsic} values of the observed $\Lo-\Lx$ relation. The fact that the observed $\Lo-\Lx$ relation is very tight is the manifestation of a common physical nature in quasars. Yet, at present, the details on the physics governing the interplay between the X--ray corona and the accretion disk is still not well understood.


\section{X--ray variability}
\label{X--ray variability}
   \begin{figure}
   \centering
   \includegraphics[width=\hsize]{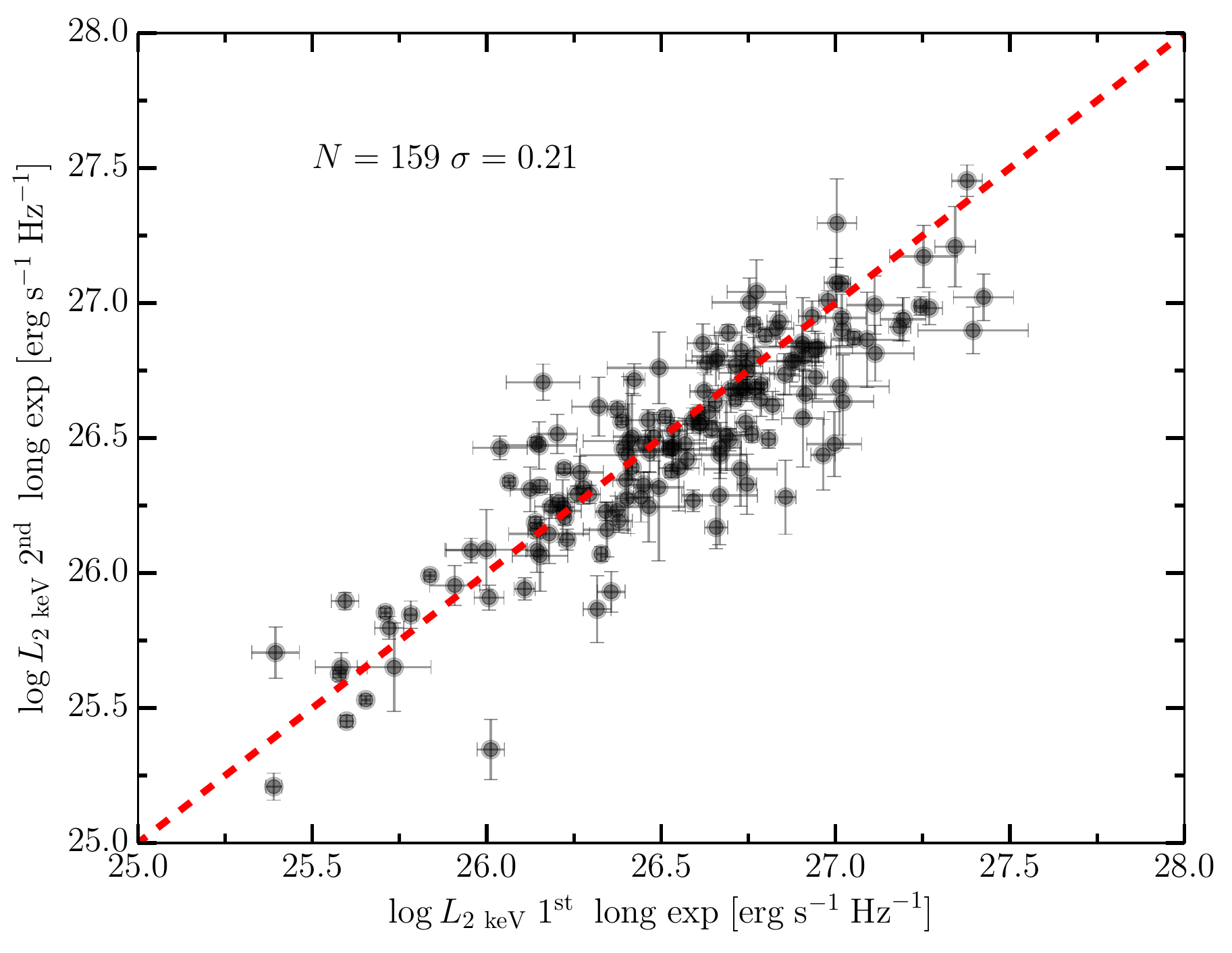}
   \includegraphics[width=\hsize]{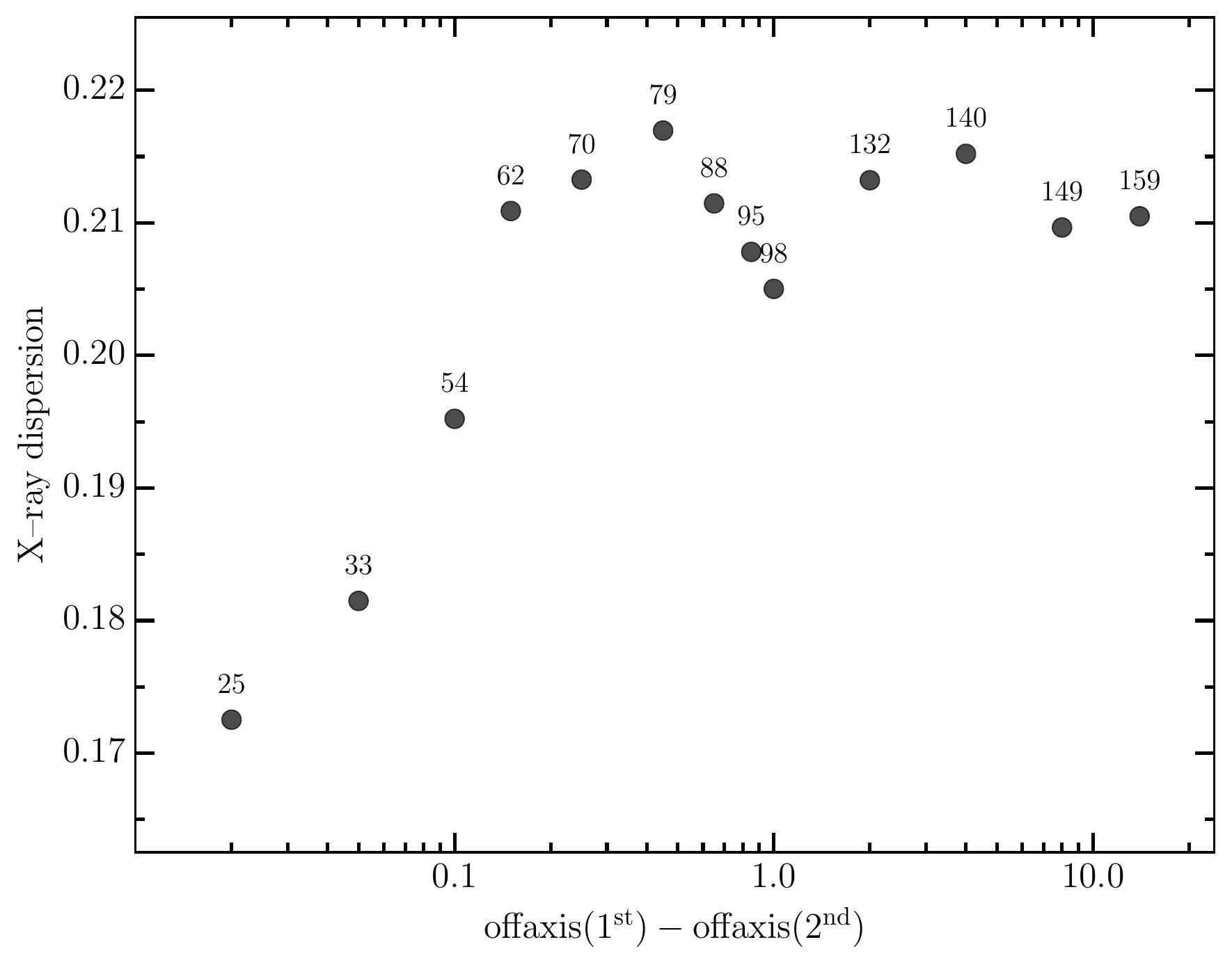}
      \caption{Upper panel: Monochromatic 2~keV luminosities of the $2^{\rm nd}$ XMM longest exposure as a function of those with the longest XMM exposure for the selected quasar sample with multiple observations. The dispersion along the one-to-one relation (red dashed line) is 0.21 dex and it roughly quantifies the extent of X--ray variability on the $\Lo-\Lx$ relationship. Lower panel: variability dispersion as a function of the offaxis difference between the two longest XMM observations. The number of objects in each offaxis bin is reported on top of each point.}
         \label{varx}
   \end{figure}
\begin{figure}
   \centering
   \includegraphics[width=\hsize]{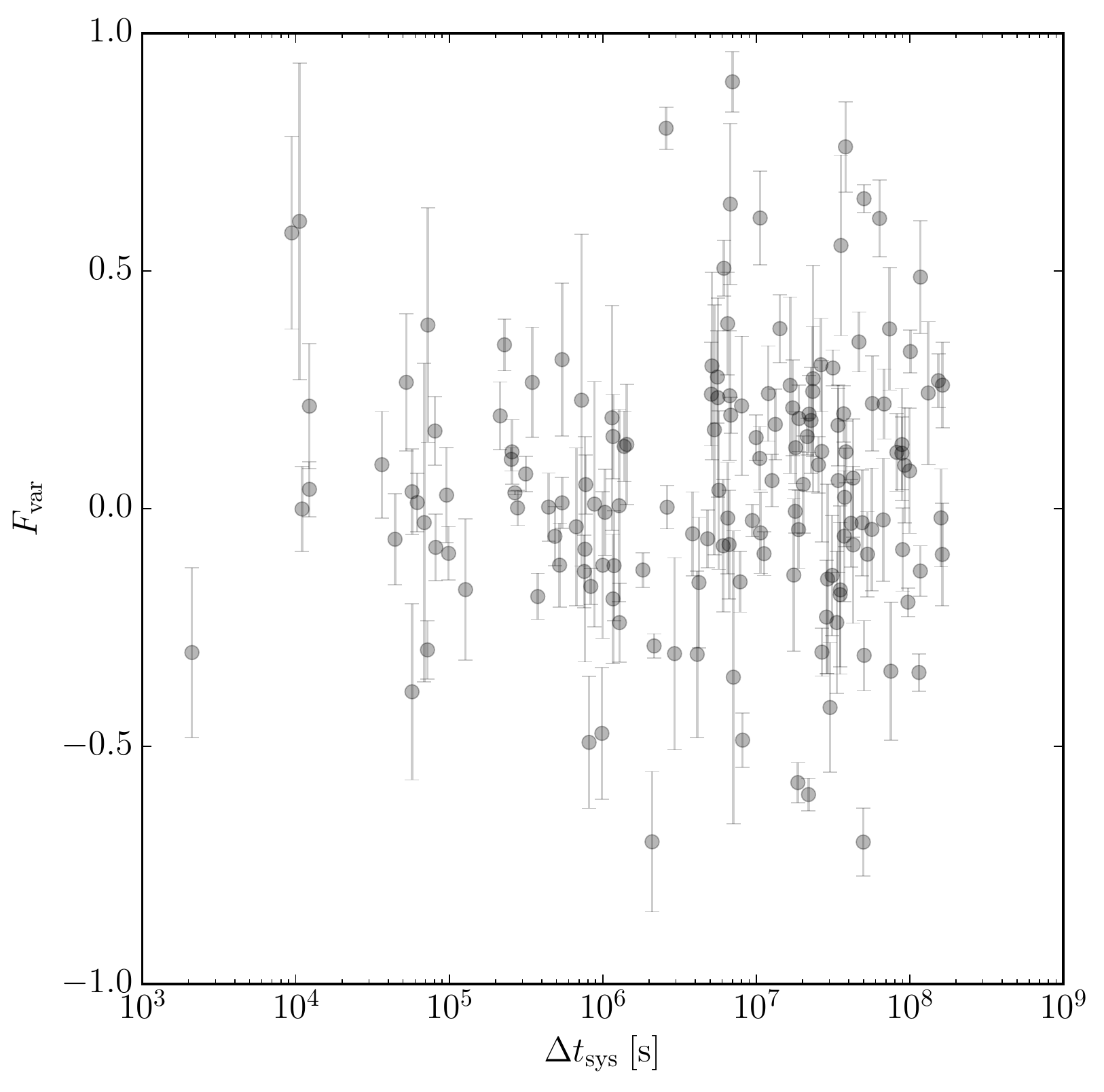}
   \includegraphics[width=\hsize]{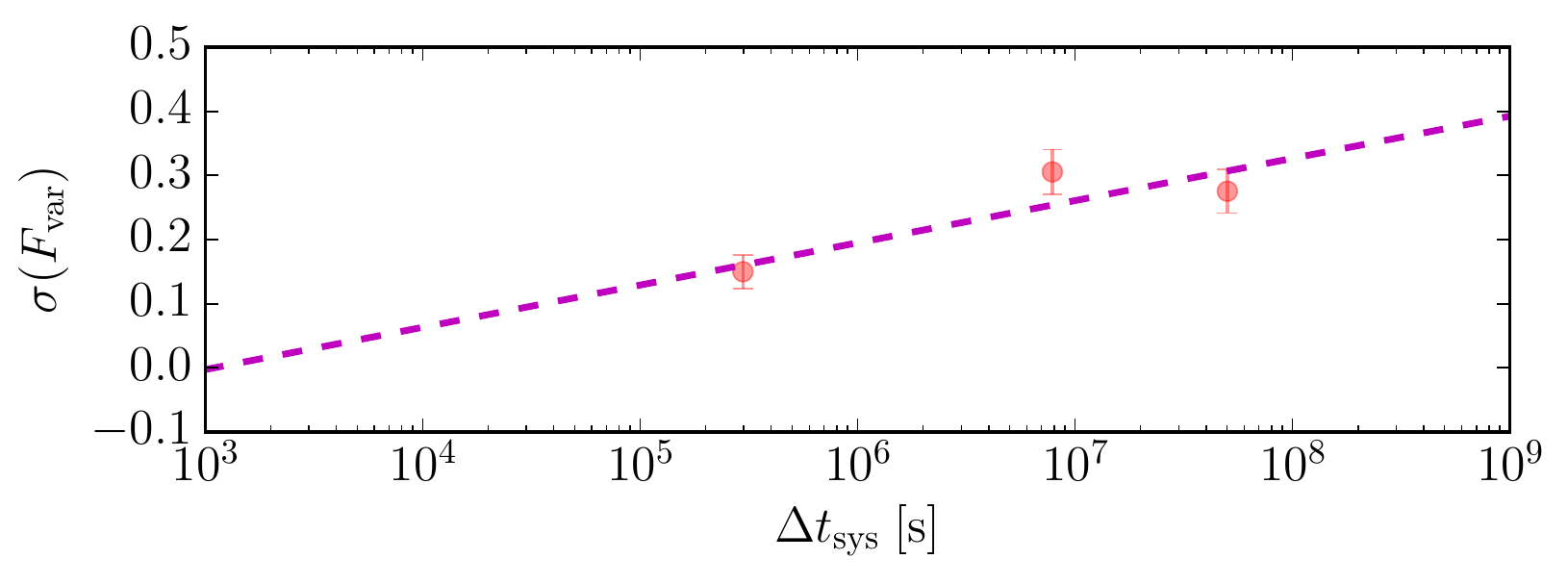}
      \caption{Upper panel: distribution of $\fvar$ as a function of the rest-frame time measurement $\deltat$. Each point is estimated taking the $1^{\rm st}$ and $2^{\rm nd}$ XMM longest exposure. Lower panel: Gaussian dispersion of $\fvar$ as a function of the rest-frame time measurement $\deltat$. The magenta dashed line is a linear fit of $\sfvar$ (see text for details).}
         \label{fvarx}
\end{figure}

One of the contributors to the scatter in the $\Lo-\Lx$ relationship can be the X--ray emission variation among different observations.
To quantify the extent of such variability, we have compared the monochromatic 2~keV luminosities of the $2^{\rm nd}$ XMM longest exposure to the ones of the longest XMM exposure. We have considered the 470 quasars with more than two XMM observations and we have applied the same selection criteria as the ones discussed in Section~\ref{The lolx relation: censored data}, leading to a sample 159 quasars. We have then estimated $\Lx$ for the second longest exposure following the same approach as for the longest one. The comparison between these two luminosities is shown in Figure~\ref{varx}. 
The dispersion along the one-to-one relation (red dashed line) is 0.21 dex. The procedure to convert counts into fluxes/luminosities introduces additional scatter, which is not due to X--ray variability alone. In fact, the observed dispersion between $\Lx(1^{\rm st})$ and $\Lx(2^{\rm nd})$ also depends on the deviation between the off-axis angles of the two observations, with the minimum difference (i.e. offaxis$(1^{\rm st})-$offaxis$(2^{\rm nd})=0.02$) having the lowest dispersion ($\sim$0.17 dex). The majority of the sources with multiple, almost on-axis, observations are indeed pointed objects where uncertainties due to flux calibration, background subtraction, and vignetting correction are almost negligible. We find that the dispersion due to X--ray variability, bad calibrations, etc, is thus on the order of $\sim$0.12 dex.

To provide another quantitative measure of the amplitude of X--ray variability in our sample we followed a similar procedure as the one described by \citet{2012ApJ...746...54G}. We computed the {\it fractional variation} ($\fvar$) as
\begin{equation}
\label{fvar}
\fvar \equiv (c_i - c_j)/(c_i + c_j),
\end{equation}
where $c_i$ and $c_j$ are the count rates for the $1^{\rm st}$ and $2^{\rm nd}$ XMM longest exposure, respectively. Each measurement of $\fvar$ between the two exposures is associated with a rest-frame time measurement $\deltat$ defined as $\deltat = (t_j - t_i)/(1+z)$, where we have taken the absolute value of $(t_j - t_i)$. 
We then assumed an intrinsic Gaussian distribution for $\fvar$ and estimated the standard deviation of this Gaussian distribution ($\sfvar$) using the likelihood method described by \citet{1988ApJ...326..680M}. We then binned $\deltat$ in three intervals of about 50 epochs each.
The upper and lower panel of Figure~\ref{fvarx} shows our results for $\fvar$ and $\sfvar$ as a function of $\deltat$ for each quasar epoch, respectively. 
Each value of $\sfvar$ is plotted at the median $\deltat$ in the considered bin. 
The dashed line represents a linear fit of $\sfvar$, which is parametrised as 
\begin{equation}
\label{sfvar}
\sfvar = (0.066\pm0.033) \log \deltat + (-0.200\pm0.215).
\end{equation}
The level of fractional variation is $\sim$20\%, which corresponds to a dispersion of $\sim$0.08 dex in log. If we consider timescales longer than 1 week ($\deltat\geq6\times10^5$s), given that such timescales represent the majority of our data set, we have that the amplitude of fractional variation is $\sim$31\% (i.e. $\sim$0.12 dex). Unfortunately, we do not have enough data to provide significant constraints on $\sfvar$ as a function of $\deltat$, hence we will consider the latter value as more representative of our sample. This value is also in agreement with the one we have estimated from the analysis of the off-axis discussed above.

As a comparison, \citet{2012ApJ...746...54G} presented a detailed analysis of the quasar X--ray variability as a function of timescale, redshift, luminosity, and optical spectral properties. Their sample consists of 264 optically selected quasars from SDSS-DR5 with at least two X--ray observations (three or more are available for 82 quasars) from the {\it Chandra} public archive. They found that $\sfvar$ is $\sim$16-17\% (corresponding to a dispersion of $\sim$0.065 dex in log) at $\deltat>5\times10^5$s, which is a rather lower value with respect to our. 
This may be partly due to the combination of higher statistics and the lower background level of their {\it Chandra} data.  

\citet{2014ApJ...781..105L} have analysed a sample of 638 AGN (340 Type 1) with XMM--{\it Newton} observations in the COSMOS field over 3.5 yr to study their long term variability. The amplitude of their fractional variation is on the order of $\sim30$\%, in close agreement with our findings. 

\begin{figure}
   \centering
   \includegraphics[width=\hsize]{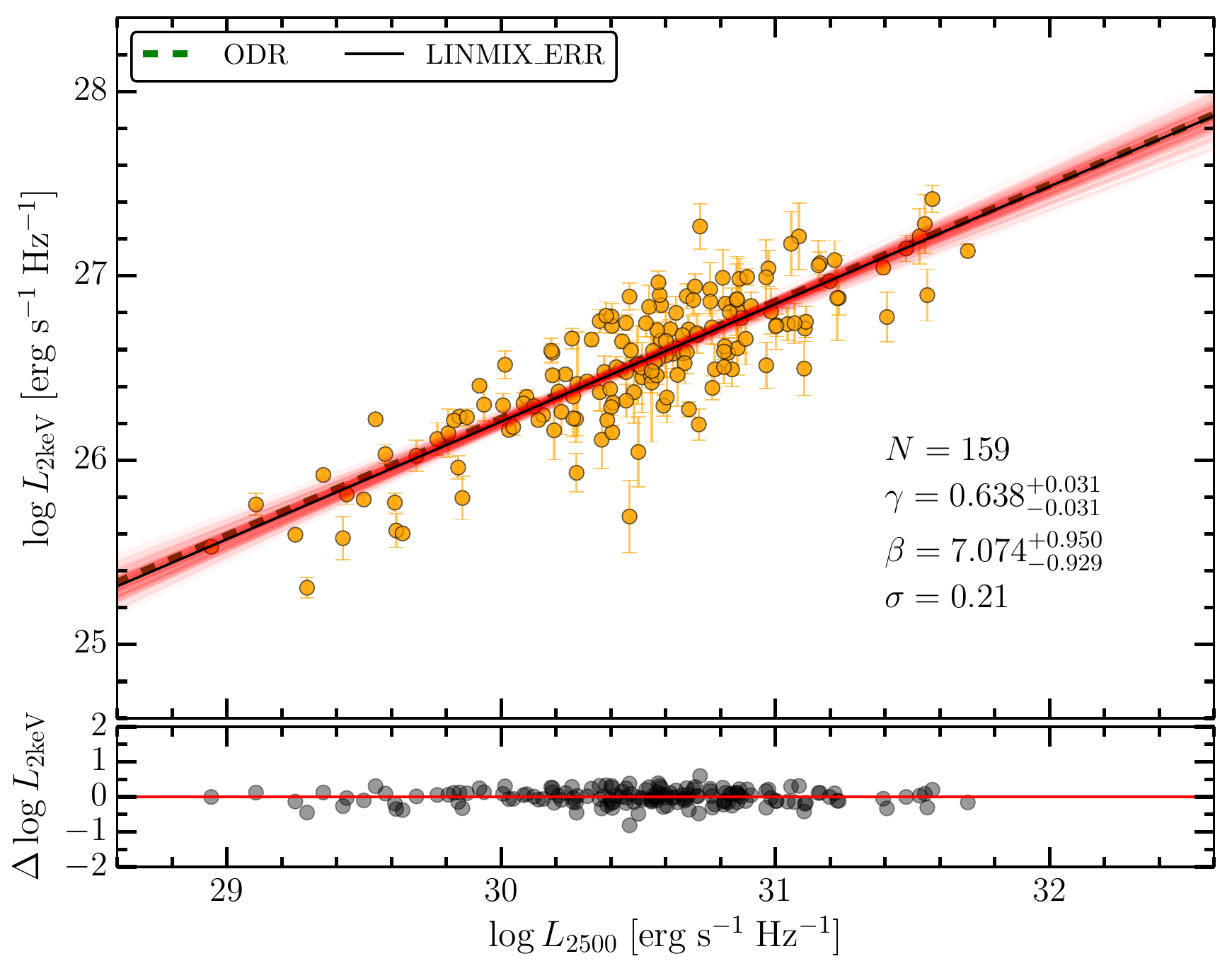}
      \caption{Keys as in Figure~\ref{lolx} for the selected quasar sample with multiple X--ray detections. See \S~\ref{X--ray variability} for details. The X--ray fluxes are estimate as the average between the $1^{\rm st}$ and $2^{\rm nd}$ XMM longest exposure.}
         \label{lolx_ave}
   \end{figure}

For the sample of 159 quasars with multiple observations, we have then calculated $\gamma$ and $\beta$ of the $\Lo-\Lx$ relation finding $\gamma=0.672\pm0.035$, $\beta=6.044^{+1.166}_{-1.075}$, and a dispersion of 0.23 dex in agreement with the best selected quasar sample. 
If we instead consider the $\Lx$ values estimated as the average between the $1^{\rm st}$ and $2^{\rm nd}$ XMM longest exposure, the dispersion on the $\Lo-\Lx$ relation reduces to 0.21 dex (see Figure~\ref{lolx_ave}), while both slope and intercept are still in agreement. 
Results are not affected even considering $\Lx$ values estimated as the average of all XMM observations. This is likely because the majority of these objects have two observations (55\%, 88/159), with solely 35 quasars with more than 3 detections.

Summarizing, we found that the amplitude of X--ray variability in the sample of 159 quasars with multiple observations is around 0.12 dex.

\section{Simultaneous observations}
\label{Simultaneous observations}
To test the influence of non-simultaneous measurements on the $\Lo-\Lx$ relationship we extracted simultaneous data from the latest release of XMM-Newton Optical Monitor Serendipitous UV Source Survey catalogue \citep[XMMSUSS2.1]{2012MNRAS.426..903P} available online.\footnote{http://heasarc.gsfc.nasa.gov/W3Browse/all/xmmomsuss.html.}
XMMSUSS2.1 includes all data to the end of 2013 and contains 7,170 observations for a total number of entries of 6,246,432 (4,329,363 unique sources) detected in one to six broad-band UV and optical filters in the Optical Monitor (XMM-OM) on board the XMM--Newton observatory. 
We matched the sample of 2090 SDSS quasars with the XMMSUSS2.1 catalogue within 3 arcsec and, by imposing the same XMM--Newton observation in the XMMSUSS2.1 catalogue, we found 1043 entries (597 unique quasars)\footnote{This has been done by comparing the OBS\_ID and OBSID flags in the 3XMM-DR5 and XMMSUSS2.1 catalogues, which uniquely identify XMM--Newton pointings.}.

To obtain the rest-frame monochromatic luminosities at 2500\AA\ we used all the available photometry compiled in the XMMSUSS2.1 catalogue along with their uncertainties. The OM set of filters covers the optical and part of the UV wavelength range: $V$, $B$, $U$, $UVW1$, $UVM2$, $UVW2$ with effective wavelengths 5430\AA, 4500\AA, 3440\AA, 2910\AA, 2310\AA, and 2120\AA. 
Galactic reddening has been taken into account: we used the selective attenuation of the stellar continuum $k(\lambda)$ taken from \citet{prevot84} with $R_{\rm V}=2.7$. Galactic extinction is estimated from \citet{schlegel98} for each object. We derived total luminosities at the rest-frame frequency of the filter. 
The data for the SED computation from mid-infrared to UV (upper limits are not considered) were then blueshifted to the rest-frame. 

The intergalactic medium (IGM) absorption by neutral hydrogen along the line of sight may introduce a hardening of the SED shape at wavelengths shorter than 1500~\AA, hence $\Lo$ could be underestimated. This effect is much stronger at $\lambda<912$\AA\ (see \citealt{2014MNRAS.438..476P}). In the SED construction we have been rather conservative and we have neglected all rest-frame data at wavelengths shorter than 1500~\AA\ ($\log\nu/{\rm Hz}>15.30$). 
In the case of two data points in-between 1500~\AA\ (but still at $\lambda<912$ \AA), we excluded the second entry at $\lambda<1500$\AA. 

In the case 2500~\AA\ were covered by no less than two data points, the $\Lo$ values are extracted from the rest-frame SEDs in the $\log \nu-\log\nu L(\nu)$ plane. 
If the SED is constructed by two data points (or more), but they do not cover 2500 \AA, luminosities are extrapolated by considering the last (first) two photometric data points. 
Uncertainties on monochromatic luminosities from interpolation (extrapolation) are computed as in equation~(\ref{uncertainties}).
In the case the SED had only one data point in the wavelength range $1500-4000$\AA, we computed $\Lo$ by employing a fixed optical slope $\gamma\sim0.61\pm0.01$ \citep{2002ApJ...565..773T,2012ApJ...752..162S,2014arXiv1408.5900S,2015MNRAS.449.4204L}. The uncertainty on the slope has been properly propagated.

We estimated monochromatic rest-frame luminosities at 2500~\AA\ for 118 quasars in the selected X--ray detected sample, which are in very good agreement with the ones we considered in our analysis although with a moderate dispersion ($\sim0.2$ dex). Such dispersion is mostly due to the different methodology used to compute the two optical--UV luminosities. Shen et al. luminosities are estimated through a continuum fit, while the OM ones still have some emission line contribution. Thus, the {\it true} optical--UV variability should be much lower, although it is not straightforward to quantify to what degree. 
We then computed slope, intercept, and dispersion (using LINMIX\_ERR) of the $\Lo-\Lx$ relation for the 118 quasars within the clean X--ray detected sample with simultaneous $\Lo$ values finding $\gamma=0.697\pm0.040$, $\beta=5.199^{+1.201}_{-1.211}$, and a dispersion of 0.25 dex. The same procedure has been done by replacing the optical luminosities with the non-simultaneous values finding $\gamma=0.694\pm0.039$, $\beta=5.333\pm1.180$, and $\sigma=0.25$ dex, which is fully consistent with the simultaneous fit.

This finding is consistent with the work done by \citet{2010A&A...519A..17V}. They have analysed simultaneous observations for 241 quasars (46 sources with multi-epoch data and 195 objects with single-epoch observations) having the X--ray/optical-UV information from the first releases of both XMM and OM catalogues (see their Section~2 for details). The observed dispersion in their simultaneous data was not significantly smaller than what previously found in non-simultaneous studies, which indicates that the ``artificial variability'' introduced by the non-simultaneity was not the main source of dispersion. 

\section{Discussion and Conclusions}
\label{Conclusions}
The observed $\Lo-\Lx$ relationship in quasars indicates that there is a good ``coupling" between the disk, emitting the primary radiation, and the hot-electron corona, emitting X--rays. 
Earlier studies have found that the scatter on this relation is $\sim0.35-0.4$ dex, which is a combination of measurement \rev{uncertainties}, variability, and intrinsic dispersion due to differences in the AGN physical properties.
We analysed in depth the various sources of the observed dispersion on the $\Lo-\Lx$ relationship for a sample of 2,153 optically selected quasars with X--ray data from the 3XMM--DR5 source catalogue.

Our study shows that, once a homogeneous quasar sample is selected, the observed dispersion on the $\Lo-\Lx$ relationship, for a sample of 159 quasars with multiple observations, is $\sim0.23$ dex considering the longest X--ray exposure.
If we instead average X--ray luminosity values, we found that the scatter further reduces to $\sim0.21$ dex. This result can be parametrised as follows
$$
  \begin{cases} 
   0.23 = \sqrt{\sigmax^2 + \sigmai^2} \\
   0.21 = \sqrt{\frac{\sigmax^2}{2} + \sigmai^2}
  \end{cases}
$$
where $\sigmax$ is the contribution of the observed X--ray variability to the dispersion, and $\sigmai$ denotes all other possible sources of scatter (e.g. uncertainties on X--ray calibrations, optical variability, intrinsic variability related to AGN physics). In other words, $\sigmai$ represent our {\it ignorance} on the observed dispersion. From the simple system above we have: $\sigmax = 0.13$ (consistent with the findings discussed in Section~\ref{X--ray variability}) and $\sigmai=0.19$. Variability in the optical is on the order of 0.05 dex (\citealt{2010ApJ...708..927K} and references therein), which gives us a residual scatter of 0.18 dex. This low scatter provides a stringent observational constraint that any future self-consistent disk--corona models must explain.
The tight $\Lo-\Lx$ relationship discussed here also allows to accurately estimate the quasar X--ray luminosity for a given optical-UV luminosity, in order to define a {\it standard} range of soft X--ray emission for {\it typical} quasars (i.e., non-BAL/RL with minimum contamination of absorption), or vice-versa to easily identify peculiar objects (e.g. X--ray weak, strong RL).

Finally, establishing a tight correlation between $\Lo$ and $\Lx$ for a statistically significant quasar sample over a wide range of redshifts and luminosities is the first step to robustly estimate cosmological parameters \citep{2015ApJ...815...33R}. 
One of the main limitation of the standard candle approach lies in the large observed dispersion of the relation ($\sim$0.3 dex). Although the quasar sample analysed here presents a remarkably tight $\Lo-\Lx$ relationship, the low number of quasars, especially at high redshift (e.g. only 14 and 1 quasar have $z>3$ within the X--ray detected sample composed by 743 and 159 sources, respectively), allows us to obtain only loose constraints on $\Omega_{\rm M}$, and $\Omega_\Lambda$. XMM--{\it Newton} and {\it Chandra} observations of larger samples of high$-z$ quasars would better sample the Hubble diagram and provide tighter constraints on cosmological parameters (Lusso et al., in preparation).

In summary, our main findings are the following. 
\begin{enumerate}
\item If we consider an optically selected quasar sample having minimum host-galaxy/reddening contamination and X--ray absorption, and reasonable X--ray S/N ratio (i.e. $>5$), the dispersion is much lower than what previously reported in the literature (0.24 dex against 0.35--0.4 dex). The slope and intercept are, overall, consistent (within their uncertainties) with $\sim0.60-0.65$ and $\sim7-8$, respectively, and they are almost independent on the {\it quality cuts} we applied. 

\item Quasars having at least two X--ray observations in the cleaned data set (159 sources) typically vary with a standard deviation of fractional variation of about 30\%.

\item Based on our analysis of the correlation in a sub-sample with multiple observations, we measure a dispersion as low as 0.21, and we estimate that the ``intrinsic'' dispersion, i.e. the dispersion not due to measurement statistical and/or systematic \rev{uncertainty}, is lower than 0.19.

\item The dispersion on the $\Lo-\Lx$ relation estimated taking simultaneous data is not significantly smaller than what we find in the non-simultaneous sample. This indicates that the ``artificial variability'' introduced by the non-simultaneity is not the main source of dispersion. 

\end{enumerate}

\begin{acknowledgements}
We thank the anonymous referee for his/her useful comments and suggestions which have improved the clarity of the paper.
We also thank Cristian Vignali for carefully reading our paper, Giorgio Lanzuisi for suggestions on quantifying X--ray variability, Fausto Vagnetti for clarifications about sample matching, Andrea Comastri and Piero Ranalli for comments on the quasar sample with X--ray upper limits.
E. L. is grateful to Keele University for the hospitality.
For all catalogue correlations we have used the Virtual Observatory software TOPCAT \citep{2005ASPC..347...29T} available online (http://www.star.bris.ac.uk/$\sim$mbt/topcat/).
This research has made use of data obtained from the 3XMM XMM--Newton serendipitous source catalogue compiled by the 10 institutes of the XMM--Newton Survey Science Centre selected by ESA. This research has made use of the XMM--OM Serendipitous Ultra-violet Source Survey, which has been created at the University College London's (UCL's) Mullard Space Science Laboratory (MSSL) on behalf of ESA and is a partner resource to the 3XMM serendipitous X--ray source catalogue. This research made use of matplotlib, a Python library for publication quality graphics \citep{2007CSE.....9...90H}. This work has been supported by the grant PRIN-INAF 2012.
\end{acknowledgements}

\bibliographystyle{apj}
\bibliography{bibl}

\begin{thebibliography}{50}
\expandafter\ifx\csname natexlab\endcsname\relax\def\natexlab#1{#1}\fi

\bibitem[{{Avni} \& {Tananbaum}(1982)}]{avnitananbaum82}
{Avni}, Y., \& {Tananbaum}, H. 1982, \apjl, 262, L17

\bibitem[{{Bechtold} {et~al.}(2003){Bechtold}, {Siemiginowska}, {Shields},
  {Czerny}, {Janiuk}, {Hamann}, {Aldcroft}, {Elvis}, \&
  {Dobrzycki}}]{bechtold03}
{Bechtold}, J., {Siemiginowska}, A., {Shields}, J., {et~al.} 2003, \apj, 588,
  119

\bibitem[{{Chanan}(1983)}]{chanan83}
{Chanan}, G.~A. 1983, \apj, 275, 45

\bibitem[{{Dadina}(2008)}]{2008A&A...485..417D}
{Dadina}, M. 2008, \aap, 485, 417

\bibitem[{{Feigelson} \& {Nelson}(1985)}]{feigelsonnelson85}
{Feigelson}, E.~D., \& {Nelson}, P.~I. 1985, \apj, 293, 192

\bibitem[{{Gibson} \& {Brandt}(2012)}]{2012ApJ...746...54G}
{Gibson}, R.~R., \& {Brandt}, W.~N. 2012, \apj, 746, 54

\bibitem[{{Gibson} {et~al.}(2009){Gibson}, {Jiang}, {Brandt}, {Hall}, {Shen},
  {Wu}, {Anderson}, {Schneider}, {Vanden Berk}, {Gallagher}, {Fan}, \&
  {York}}]{2009ApJ...692..758G}
{Gibson}, R.~R., {Jiang}, L., {Brandt}, W.~N., {et~al.} 2009, \apj, 692, 758

\bibitem[{{Haardt} \& {Maraschi}(1991)}]{1991ApJ...380L..51H}
{Haardt}, F., \& {Maraschi}, L. 1991, \apjl, 380, L51

\bibitem[{{Haardt} \& {Maraschi}(1993)}]{1993ApJ...413..507H}
---. 1993, \apj, 413, 507

\bibitem[{{Hewett} \& {Wild}(2010)}]{2010MNRAS.405.2302H}
{Hewett}, P.~C., \& {Wild}, V. 2010, \mnras, 405, 2302

\bibitem[{{Hunter}(2007)}]{2007CSE.....9...90H}
{Hunter}, J.~D. 2007, Computing in Science and Engineering, 9, 90

\bibitem[{{Isobe} {et~al.}(1990){Isobe}, {Feigelson}, {Akritas}, \&
  {Babu}}]{isobe90}
{Isobe}, T., {Feigelson}, E.~D., {Akritas}, M.~G., \& {Babu}, G.~J. 1990, \apj,
  364, 104

\bibitem[{{Isobe} {et~al.}(1986){Isobe}, {Feigelson}, \& {Nelson}}]{isobe86}
{Isobe}, T., {Feigelson}, E.~D., \& {Nelson}, P.~I. 1986, \apj, 306, 490

\bibitem[{{Just} {et~al.}(2007){Just}, {Brandt}, {Shemmer}, {Steffen},
  {Schneider}, {Chartas}, \& {Garmire}}]{just07}
{Just}, D.~W., {Brandt}, W.~N., {Shemmer}, O., {et~al.} 2007, \apj, 665, 1004

\bibitem[{{Kellermann} {et~al.}(1989){Kellermann}, {Sramek}, {Schmidt},
  {Shaffer}, \& {Green}}]{kellermann89}
{Kellermann}, K.~I., {Sramek}, R., {Schmidt}, M., {Shaffer}, D.~B., \& {Green},
  R. 1989, \aj, 98, 1195

\bibitem[{{Kelly}(2007)}]{2007ApJ...665.1489K}
{Kelly}, B.~C. 2007, \apj, 665, 1489

\bibitem[{{Komatsu} {et~al.}(2009){Komatsu}, {Dunkley}, {Nolta}, {Bennett},
  {Gold}, {Hinshaw}, {Jarosik}, {Larson}, {Limon}, {Page}, {Spergel},
  {Halpern}, {Hill}, {Kogut}, {Meyer}, {Tucker}, {Weiland}, {Wollack}, \&
  {Wright}}]{komatsu09}
{Komatsu}, E., {Dunkley}, J., {Nolta}, M.~R., {et~al.} 2009, \apjs, 180, 330

\bibitem[{{Koz{\l}owski} {et~al.}(2010){Koz{\l}owski}, {Kochanek}, {Udalski},
  {Wyrzykowski}, {Soszy{\'n}ski}, {Szyma{\'n}ski}, {Kubiak}, {Pietrzy{\'n}ski},
  {Szewczyk}, {Ulaczyk}, {Poleski}, \& {OGLE
  Collaboration}}]{2010ApJ...708..927K}
{Koz{\l}owski}, S., {Kochanek}, C.~S., {Udalski}, A., {et~al.} 2010, \apj, 708,
  927

\bibitem[{{Kriss} \& {Canizares}(1985)}]{krisscanizares85}
{Kriss}, G.~A., \& {Canizares}, C.~R. 1985, \apj, 297, 177

\bibitem[{{La Franca} {et~al.}(1995){La Franca}, {Franceschini}, {Cristiani},
  \& {Vio}}]{lafranca95}
{La Franca}, F., {Franceschini}, A., {Cristiani}, S., \& {Vio}, R. 1995, \aap,
  299, 19

\bibitem[{{Lanzuisi} {et~al.}(2014){Lanzuisi}, {Ponti}, {Salvato}, {Hasinger},
  {Cappelluti}, {Bongiorno}, {Brusa}, {Lusso}, {Nandra}, {Merloni},
  {Silverman}, {Trump}, {Vignali}, {Comastri}, {Gilli}, {Schramm},
  {Steinhardt}, {Sanders}, {Kartaltepe}, {Rosario}, \&
  {Trakhtenbrot}}]{2014ApJ...781..105L}
{Lanzuisi}, G., {Ponti}, G., {Salvato}, M., {et~al.} 2014, \apj, 781, 105

\bibitem[{{Lavalley} {et~al.}(1992){Lavalley}, {Isobe}, \&
  {Feigelson}}]{lavalley92}
{Lavalley}, M., {Isobe}, T., \& {Feigelson}, E. 1992, in Astronomical Society
  of the Pacific Conference Series, Vol.~25, Astronomical Data Analysis
  Software and Systems I, ed. D.~M. {Worrall}, C.~{Biemesderfer}, \&
  J.~{Barnes}, 245

\bibitem[{{Lusso} {et~al.}(2015){Lusso}, {Worseck}, {Hennawi}, {Prochaska},
  {Vignali}, {Stern}, \& {O'Meara}}]{2015MNRAS.449.4204L}
{Lusso}, E., {Worseck}, G., {Hennawi}, J.~F., {et~al.} 2015, \mnras, 449, 4204

\bibitem[{{Lusso} {et~al.}(2010)}]{2010A&A...512A..34L}
{Lusso}, E., {et~al.} 2010, \aap, 512, A34

\bibitem[{{Maccacaro} {et~al.}(1988){Maccacaro}, {Gioia}, {Wolter}, {Zamorani},
  \& {Stocke}}]{1988ApJ...326..680M}
{Maccacaro}, T., {Gioia}, I.~M., {Wolter}, A., {Zamorani}, G., \& {Stocke},
  J.~T. 1988, \apj, 326, 680

\bibitem[{{Page} {et~al.}(2012){Page}, {Brindle}, {Talavera}, {Still}, {Rosen},
  {Yershov}, {Ziaeepour}, {Mason}, {Cropper}, {Breeveld}, {Loiseau}, {Mignani},
  {Smith}, \& {Murdin}}]{2012MNRAS.426..903P}
{Page}, M.~J., {Brindle}, C., {Talavera}, A., {et~al.} 2012, \mnras, 426, 903

\bibitem[{{Prevot} {et~al.}(1984){Prevot}, {Lequeux}, {Prevot}, {Maurice}, \&
  {Rocca-Volmerange}}]{prevot84}
{Prevot}, M.~L., {Lequeux}, J., {Prevot}, L., {Maurice}, E., \&
  {Rocca-Volmerange}, B. 1984, \aap, 132, 389

\bibitem[{{Prochaska} {et~al.}(2014){Prochaska}, {Madau}, {O'Meara}, \&
  {Fumagalli}}]{2014MNRAS.438..476P}
{Prochaska}, J.~X., {Madau}, P., {O'Meara}, J.~M., \& {Fumagalli}, M. 2014,
  \mnras, 438, 476

\bibitem[{{Richards} {et~al.}(2006)}]{2006AJ....131.2766R}
{Richards}, G.~T., {et~al.} 2006, \aj, 131, 2766

\bibitem[{{Risaliti} \& {Lusso}(2015)}]{2015ApJ...815...33R}
{Risaliti}, G., \& {Lusso}, E. 2015, \apj, 815, 33

\bibitem[{{Rosen} {et~al.}(2015){Rosen}, {Webb}, {Watson}, {Ballet}, {Barret},
  {Braito}, {Carrera}, {Ceballos}, {Coriat}, {Della Ceca}, {Denkinson},
  {Esquej}, {Farrell}, {Freyberg}, {Gris{\'e}}, {Guillout}, {Heil},
  {Law-Green}, {Lamer}, {Lin}, {Martino}, {Michel}, {Motch}, {Nebot
  Gomez-Moran}, {Page}, {Page}, {Page}, {Pakull}, {Pye}, {Read}, {Rodriguez},
  {Sakano}, {Saxton}, {Schwope}, {Scott}, {Sturm}, {Traulsen}, {Yershov}, \&
  {Zolotukhin}}]{2015arXiv150407051R}
{Rosen}, S.~R., {Webb}, N.~A., {Watson}, M.~G., {et~al.} 2015, ArXiv e-prints

\bibitem[{{Schlegel} {et~al.}(1998){Schlegel}, {Finkbeiner}, \&
  {Davis}}]{schlegel98}
{Schlegel}, D.~J., {Finkbeiner}, D.~P., \& {Davis}, M. 1998, \apj, 500, 525

\bibitem[{{Shen} {et~al.}(2011){Shen}, {Richards}, {Strauss}, {Hall},
  {Schneider}, {Snedden}, {Bizyaev}, {Brewington}, {Malanushenko},
  {Malanushenko}, {Oravetz}, {Pan}, \& {Simmons}}]{2011ApJS..194...45S}
{Shen}, Y., {Richards}, G.~T., {Strauss}, M.~A., {et~al.} 2011, \apjs, 194, 45

\bibitem[{{Shull} {et~al.}(2012){Shull}, {Stevans}, \&
  {Danforth}}]{2012ApJ...752..162S}
{Shull}, J.~M., {Stevans}, M., \& {Danforth}, C.~W. 2012, \apj, 752, 162

\bibitem[{{Steffen} {et~al.}(2006){Steffen}, {Strateva}, {Brandt}, {Alexander},
  {Koekemoer}, {Lehmer}, {Schneider}, \& {Vignali}}]{steffen06}
{Steffen}, A.~T., {Strateva}, I., {Brandt}, W.~N., {et~al.} 2006, \aj, 131,
  2826

\bibitem[{{Stevans} {et~al.}(2014){Stevans}, {Shull}, {Danforth}, \&
  {Tilton}}]{2014arXiv1408.5900S}
{Stevans}, M.~L., {Shull}, J.~M., {Danforth}, C.~W., \& {Tilton}, E.~M. 2014,
  ArXiv e-prints

\bibitem[{{Strateva} {et~al.}(2005){Strateva}, {Brandt}, {Schneider}, {Vanden
  Berk}, \& {Vignali}}]{strateva05}
{Strateva}, I.~V., {Brandt}, W.~N., {Schneider}, D.~P., {Vanden Berk}, D.~G.,
  \& {Vignali}, C. 2005, \aj, 130, 387

\bibitem[{{Tananbaum} {et~al.}(1979){Tananbaum}, {Avni}, {Branduardi}, {Elvis},
  {Fabbiano}, {Feigelson}, {Giacconi}, {Henry}, {Pye}, {Soltan}, \&
  {Zamorani}}]{avnitananbaum79}
{Tananbaum}, H., {Avni}, Y., {Branduardi}, G., {et~al.} 1979, \apjl, 234, L9

\bibitem[{{Tang} {et~al.}(2007){Tang}, {Zhang}, \&
  {Hopkins}}]{2007MNRAS.377.1113T}
{Tang}, S.~M., {Zhang}, S.~N., \& {Hopkins}, P.~F. 2007, \mnras, 377, 1113

\bibitem[{{Taylor}(2005)}]{2005ASPC..347...29T}
{Taylor}, M.~B. 2005, in Astronomical Society of the Pacific Conference Series,
  Vol. 347, Astronomical Data Analysis Software and Systems XIV, ed.
  P.~{Shopbell}, M.~{Britton}, \& R.~{Ebert}, 29

\bibitem[{{Telfer} {et~al.}(2002){Telfer}, {Zheng}, {Kriss}, \&
  {Davidsen}}]{2002ApJ...565..773T}
{Telfer}, R.~C., {Zheng}, W., {Kriss}, G.~A., \& {Davidsen}, A.~F. 2002, \apj,
  565, 773

\bibitem[{{Vagnetti} {et~al.}(2010){Vagnetti}, {Turriziani}, {Trevese}, \&
  {Antonucci}}]{2010A&A...519A..17V}
{Vagnetti}, F., {Turriziani}, S., {Trevese}, D., \& {Antonucci}, M. 2010, \aap,
  519, A17

\bibitem[{{Vasudevan} \& {Fabian}(2009)}]{2009MNRAS.392.1124V}
{Vasudevan}, R.~V., \& {Fabian}, A.~C. 2009, \mnras, 392, 1124

\bibitem[{{Vignali} {et~al.}(2003){Vignali}, {Brandt}, \&
  {Schneider}}]{vignali03}
{Vignali}, C., {Brandt}, W.~N., \& {Schneider}, D.~P. 2003, \aj, 125, 433

\bibitem[{{Watson} {et~al.}(2001){Watson}, {Augu{\`e}res}, {Ballet}, {Barcons},
  {Barret}, {Boer}, {Boller}, {Bromage}, {Brunner}, {Carrera}, {Cropper},
  {Denby}, {Ehle}, {Elvis}, {Fabian}, {Freyberg}, {Guillout}, {Hameury},
  {Hasinger}, {Hinshaw}, {Maccacaro}, {Mason}, {McMahon}, {Michel}, {Mirioni},
  {Mittaz}, {Motch}, {Olive}, {Osborne}, {Page}, {Pakull}, {Perry}, {Pierre},
  {Pietsch}, {Pye}, {Read}, {Roberts}, {Rosen}, {Sauvageot}, {Schwope},
  {Sekiguchi}, {Stewart}, {Stewart}, {Valtchanov}, {Ward}, {Warwick}, {West},
  {White}, \& {Worrall}}]{2001A&A...365L..51W}
{Watson}, M.~G., {Augu{\`e}res}, J.-L., {Ballet}, J., {et~al.} 2001, \aap, 365,
  L51

\bibitem[{{Watson} {et~al.}(2009){Watson}, {Schr{\"o}der}, {Fyfe}, {Page},
  {Lamer}, {Mateos}, {Pye}, {Sakano}, {Rosen}, {Ballet}, {Barcons}, {Barret},
  {Boller}, {Brunner}, {Brusa}, {Caccianiga}, {Carrera}, {Ceballos}, {Della
  Ceca}, {Denby}, {Denkinson}, {Dupuy}, {Farrell}, {Fraschetti}, {Freyberg},
  {Guillout}, {Hambaryan}, {Maccacaro}, {Mathiesen}, {McMahon}, {Michel},
  {Motch}, {Osborne}, {Page}, {Pakull}, {Pietsch}, {Saxton}, {Schwope},
  {Severgnini}, {Simpson}, {Sironi}, {Stewart}, {Stewart}, {Stobbart}, {Tedds},
  {Warwick}, {Webb}, {West}, {Worrall}, \& {Yuan}}]{2009A&A...493..339W}
{Watson}, M.~G., {Schr{\"o}der}, A.~C., {Fyfe}, D., {et~al.} 2009, \aap, 493,
  339

\bibitem[{{Wilkes} {et~al.}(1994){Wilkes}, {Tananbaum}, {Worrall}, {Avni},
  {Oey}, \& {Flanagan}}]{wilkes94}
{Wilkes}, B.~J., {Tananbaum}, H., {Worrall}, D.~M., {et~al.} 1994, \apjs, 92,
  53

\bibitem[{{Young} {et~al.}(2009){Young}, {Elvis}, \&
  {Risaliti}}]{2009ApJS..183...17Y}
{Young}, M., {Elvis}, M., \& {Risaliti}, G. 2009, \apjs, 183, 17

\bibitem[{{Young} {et~al.}(2010){Young}, {Elvis}, \&
  {Risaliti}}]{2010ApJ...708.1388Y}
---. 2010, \apj, 708, 1388

\bibitem[{{Yuan} {et~al.}(1998){Yuan}, {Siebert}, \&
  {Brinkmann}}]{yuansiebertbrink98}
{Yuan}, W., {Siebert}, J., \& {Brinkmann}, W. 1998, \aap, 334, 498

\end{thebibliography}

\appendix
\section{appendix A\\ The flux-limit bias}
\label{The flux-limit bias}
   \begin{figure}
   \centering
   \includegraphics[width=0.48\textwidth]{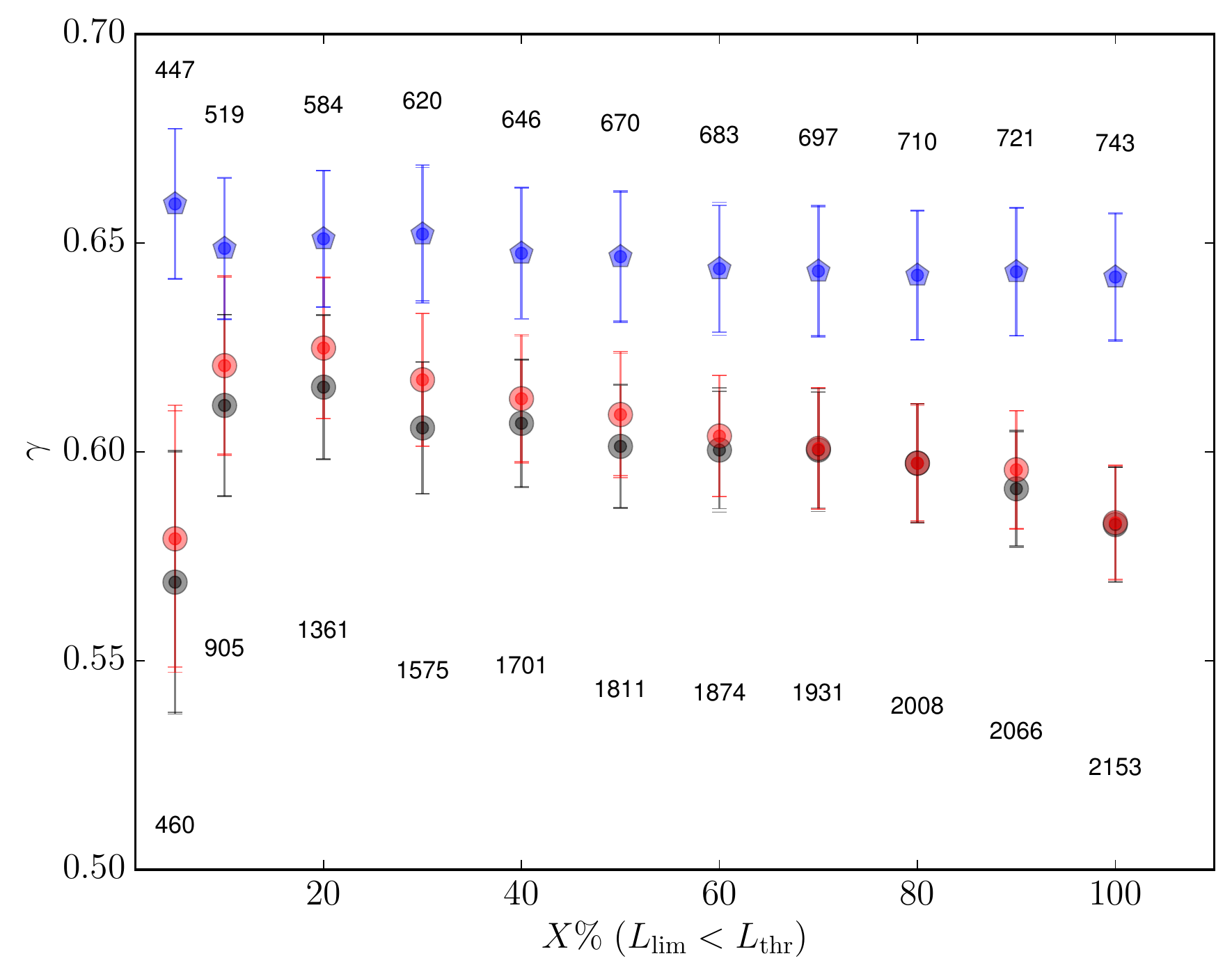}
   \includegraphics[width=0.48\textwidth]{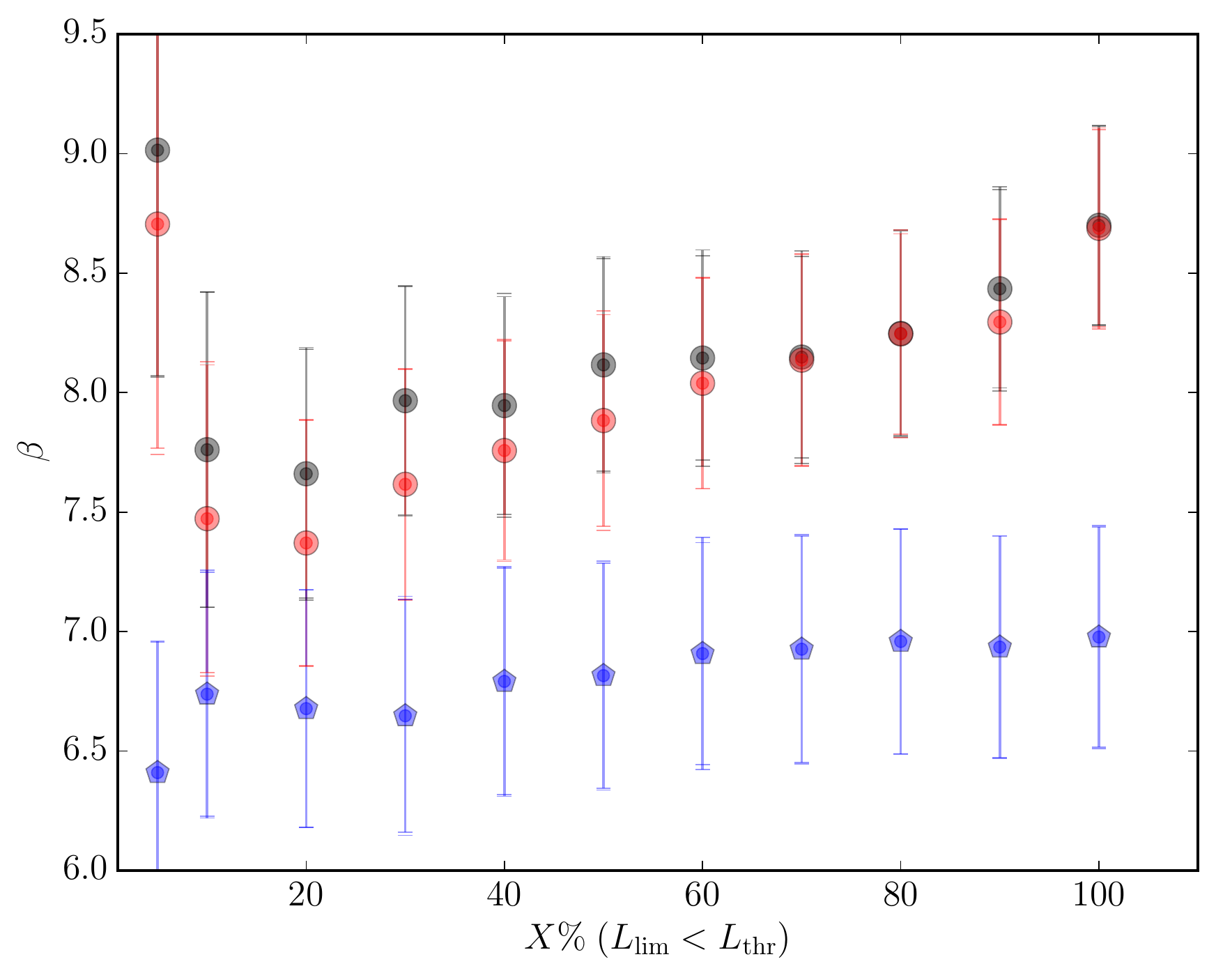}
      \caption{The distribution of the slope $\gamma$ and intercept $\beta$ with their uncertainties (values obtained by using the LINMIX\_ERR algorithm) as a function of percentage cut (see text for details). The black and red points are estimated with $\gamma'$ equal to 0.6 and 0.7, respectively. The blue points represent the trend for the selected quasars only with $\gamma'$ equal to 0.6. The number of objects that fulfil the selection in each slice is reported on top of the plotted values.} 
         \label{bias_lim}
   \end{figure}
As already pointed out in Section~\ref{Statistical analysis}, flux limited samples may be biased, and thus one needs to find an alternative method to obtain an (almost) unbiased sample even without the inclusion of censored data.
To do so, we have sliced the main quasar sample of X--ray detected quasars (2,155 objects) including only those sources that have a minimum X--ray luminosity ($\lmin$, as estimated from the $\fmin$ at 2 keV, see \S~\ref{X--ray luminosities}) below a threshold defined as follows
\begin{equation}
\label{lthr}
\log \Lthr = \gamma' \log \Lo + \beta',
\end{equation}
where $\gamma'$ is assumed to be 0.6 and $\beta'$ is a variable normalisation. 
The normalisation $\beta'$ is defined so that the source fraction enclosed ($\Lx<\Lthr$) is in the range from 5\% to 100\%.
We then included the objects in the sample only if $\lmin$ is below the detection limit given by Equation~(\ref{lthr}), thus regardless of the {\em observed} value of $\Lx$. We checked the effects of this cut varying the rejection fraction. The slope $\gamma$ and intercept $\beta$ (along with their uncertainties) of the $\Lo-\Lx$ relation for each slice are estimated by using the LINMIX\_ERR algorithm. 
We have also repeated the same approach with $\gamma'=0.7$. Figure~\ref{bias_lim} shows the results of this experiment. Black and red points represent the findings for $\gamma'=0.6$ and 0.7, respectively. Blue points are the outcome considering $\gamma'=0.6$ where we applied the quality cuts discusses in Section~\ref{Statistical analysis}.
If the flux-limit bias were strongly affecting the slope measurements in the X--ray detected quasar sample, we would have expected a flatter slope ($\gamma=0.7-0.8$) for higher cuts ($X\%\lesssim20$). In other words, a decreasing of $\gamma$ as a function of $X\%$. On the other hand, the intercept $\beta$ must increase as a function of $X\%$. 
The regression parameters are, instead, not strongly dependent (within their uncertainties) either on the choice of $\gamma'$ in Equation~(\ref{lthr}) or on the threshold. The $\gamma$ and $\beta$ values resulting from the slicing of the clean quasar sample are even more flatter than for the whole sample. Meaning that, even if the flux-limit bias may still be present, is not altering our findings significantly (both $\gamma$ and $\beta$ are statistically different at less than $2\sigma$). 
This experiment ensure that our investigation of the luminosity relation can be performed even if only detections are considered.

\end{document}